\documentclass[11pt,a4paper]{article}
\usepackage{jheppub}
\usepackage{amsmath,amstext,amssymb,amscd}
\usepackage{graphicx}
\usepackage{feynmp}

\def\N4{${\cal N} = 4$}

\def\Yf{{\rm Y}}

\title{The Bethe Roots of Regge Cuts in \\[3mm] Strongly Coupled $\mathbf{{\cal{N}}= 4}$ SYM Theory}
\subheader{DESY 14-208}
\author[a]{J.\ Bartels,}
\author[b]{V.\ Schomerus}
\author[b,c]{and M.\ Sprenger}

\affiliation[a]{II.\ Institute for Theoretical Physics, Hamburg University,\\Luruper Chaussee 149, 22671 Hamburg, Germany}
\affiliation[b]{DESY Hamburg, Theory Group,\\Notkestra\ss e 85, 22607 Hamburg, Germany}
\affiliation[c]{Institute for Theoretical Physics, ETH Z\"urich,\\Wolfgang-Pauli-Strasse 27, 8093 Z\"urich, Switzerland}

\emailAdd{joachim.bartels@desy.de}
\emailAdd{volker.schomerus@desy.de}
\emailAdd{sprengerm@itp.phys.ethz.ch}

\abstract{We describe a general algorithm for the computation of the remainder function
for $n$-gluon scattering in multi-Regge kinematics for strongly coupled planar
$\mathcal{N}=4$ super Yang-Mills theory. This regime is accessible through the infrared
physics of an auxiliary quantum integrable system describing strings in AdS$_5
\times$S$^5$. Explicit formulas are presented for $n=6$ and $n=7$ external
gluons. Our results are consistent with expectations from perturbative
gauge theory.
This paper comprises the technical details for the results announced in \cite{Bartels:2014ppa}.}


\begin{document}

\newcommand{\Y}[3]{\mathrm{Y}_{#1,#2}^{\left[#3\right]}}
\newcommand{\Ynt}[3]{\mathit{Y}_{#1,#2}^{\left[#3\right]}}
\newcommand{\U}[2]{\frac{\Ynt{2}{#1}{#2}}{1+\Ynt{2}{#1}{#2}}}
\newcommand{\Yint}[2]{\left(1+\tilde{Y}_{#1,#2}(\theta')\right)}

\maketitle

\section{Introduction}

One of the major goals of modern theoretical physics is to construct
the exact S-matrix of a four-dimensional interacting quantum field theory.
It is believed that $\mathcal{N}=4$ super Yang-Mills (SYM) theory provides the
simplest example for which this task may be achieved, at least in the
planar limit. The first conjectured expression for
gluon scattering amplitudes in this theory, known as the Bern-Dixon-Smirnov
(BDS) formula \cite{Bern:2005iz}, was shown to be incomplete for $n\geq 6$ external gluons
beyond one-loop order \cite{Alday:2007he,Bartels:2008ce,Bartels:2008sc}.
The corrections to the BDS formula are captured by the so-called remainder function.
For the maximally helicity violating (MHV) configuration the remainder function is
known up to four loops \cite{Goncharov:2010jf, Dixon:2013eka, Dixon:2014voa} for
six gluons and up to two loops \cite{CaronHuot:2011ky, Golden:2014xqf} for seven gluons.
\par
While the construction of the leading loop corrections to the BDS
formula is a remarkable achievement which is based on beautiful and
non-trivial mathematical concepts, these expressions are still a
long way from an all-loop result (although first all-loop proposals
have started to emerge recently \cite{Basso:2013vsa,Basso:2013aha,
Basso:2014koa,Basso:2014nra}).
On the other hand, all-loop expressions do exist for the anomalous
dimensions of local operators in $\mathcal{N}=4$ super Yang-Mills
theory \cite{Beisert:2010jr, Gromov:2013pga, Gromov:2014caa}. In that case, the initial progress
from perturbative field theory computations \cite{Minahan:2002ve, Beisert:2003tq, Beisert:2003jj} was soon complemented
by string theoretic calculations \cite{Kazakov:2004qf} which capture the
behavior at strong coupling via the AdS/CFT correspondence
\cite{Maldacena:1997re, Witten:1998qj, Gubser:1998bc}. It were these investigations at strong
coupling, such as \cite{Arutyunov:2004vx,Beisert:2005cw, Hernandez:2006tk, Freyhult:2006vr,Beisert:2006ib}, that brought the breakthrough and paved 
the way for the first all-loop expressions of anomalous dimensions 
\cite{Beisert:2006ez}.\par

Given the way things developed for the anomalous dimensions it seems
worthwhile to invest more effort into the study of scattering
amplitudes in strongly coupled $\mathcal{N}=4$ super Yang-Mills theory.
All computations in this regime are based on the work of Alday and
Maldacena \cite{Alday:2007hr} who propose that scattering amplitudes
in strongly coupled $\mathcal{N}=4$ super Yang-Mills theory are given by the
area of a minimal surface in AdS$_5$ whose boundary is fixed
to a piecewise light-like Wilson loop on the boundary of AdS$_5$.
In a series of papers \cite{Alday:2009yn, Alday:2009dv,Alday:2010vh}
it is shown that this minimal area problem can be
reformulated through a set of coupled non-linear integral
equations, the so-called Y-system. These take a form which is
familiar from the theory of one-dimensional quantum
integrable models. They describe a system of
excitations with certain masses and chemical potentials which live on a circle and
which interact through an integrable $2 \mapsto 2$ scattering.
The interaction is engineered in such a way that the free energy of this system exactly reproduces the area of the minimal surface and thereby elegantly computes the amplitude.\par

In general, scattering amplitudes of four-dimensional quantum field theories
are complicated functions of the kinematical invariants and the
coupling constants. While the ultimate goal is to determine the
full dependence on all of these variables, it might pay off to consider special kinematical limits at first.
There are several choices which are being considered.
These include, for example, two-dimensional kinematics \cite{Heslop:2010kq, Goddard:2012cx, Torres:2013vba, Toledo:2014koa}, the limit in which the boundary polygon becomes regular \cite{Alday:2010vh, Hatsuda:2010vr, Hatsuda:2011ke, Hatsuda:2011jn, Hatsuda:2012pb} or the OPE limit of \cite{Alday:2010ku, Gaiotto:2010fk, Gaiotto:2011dt, Sever:2011da, Sever:2012qp, Basso:2013vsa, Basso:2013aha, Basso:2014koa, Basso:2014jfa}. In this
work we shall study a well-known kinematical limit that has a
long history in gauge theory, the multi-Regge limit.
\par

The multi-Regge limit has received much attention for two important
reasons. On the one hand, it describes real scattering events at high energies which are observed at particle colliders.
On the other hand, investigations in QCD revealed a second remarkable
feature: It turns out that in the multi-Regge limit
the quantities governing the scattering amplitude are controlled by the spectrum of an integrable spin
chain \cite{Lipatov:1993yb, Faddeev:1994zg}. For $\mathcal{N}=4$ super Yang-Mills theory, these quantities
are by now known to third order in perturbation theory \cite{Bartels:2008ce,Bartels:2008sc, Lipatov:2010ad, Fadin:2011we, Dixon:2012yy, Dixon:2014voa} and for the six-gluon case an all-loop conjecture was recently put forward in \cite{Basso:2014pla}. Therefore, the situation somewhat mirrors the early development in the case of anomalous dimensions where
the first few orders in perturbation theory were also controlled with the help of integrability.\par
Given this state of affairs it seems very natural to ask what the multi-Regge limit corresponds to on the strong coupling side.
In \cite{Bartels:2010ej, Bartels:2012gq} we show that the multi-Regge limit of gauge theory corresponds to the infrared, or low
temperature, limit of the Y-system. In such a low temperature limit,
one can neglect the integral terms in the non-linear integral
equations. What remains is a system of algebraic Bethe Ansatz
equations which can be solved quite easily. The aim of this work
is to explain these Bethe Ansatz equations and to explain
how to construct the remainder functions from solutions of these
equations. In order to illustrate the general algorithm we will
then calculate the remainder function in multi-Regge kinematics
for six and seven gluons. Our results are in agreement with features
that are expected from the perturbative evaluation in the
leading logarithmic approximation.

\section{Multi-Regge kinematics}
\label{sec:kinematics}

The aim of this section is to introduce the relevant kinematic
invariants and to discuss their behavior in the multi-Regge limit.
As is well known, the $n$-gluon remainder function depends on $3n-15$
cross ratios. For our discussion of $2 \to n-2$ production
amplitudes, we shall need a particular set of such cross ratios.
This is discussed in the first subsection. We then turn to the
multi-Regge limit and describe how our basis of cross ratios behaves
as we go to high energies. The multi-Regge limit is described by
the neighborhood of a particular point in the space of cross
ratios. In order to exhibit interesting multi-Regge behavior,
we must take the limit in various regions. These are discussed
in the third subsection.

\subsection{Kinematic variables}

A scattering process of $n$ massless gluons is parametrized by kinematic invariants
which can be built from products $p_i\cdot p_j$. After taking the on-shell conditions $p_i^2=0$ and momentum conservation into account, we count $D_n = \frac{1}{2}(n-1)
(n-2) - 1$ such product invariants. But these are not all independent. In fact, they
are still subject to Gram determinant relations. There are $g_n = \frac{1}{2}(n-4)
(n-5)$ such relations which arise from the fact that at most four vectors can be
linearly independent in four dimensions and are obtained from the various vanishing $5\times 5$ subdeterminants of the $(n-1) \times (n-1)$ matrix $P_{ij} =
(p_i\cdot p_j)$, for details see appendix \ref{sec:gram}. This leaves us with
$D_n-g_n = 3n-10$ independent Mandelstam invariants which parametrize our
scattering problem.\par
\begin{figure}[t]
  \centering
  \includegraphics[scale=.7]{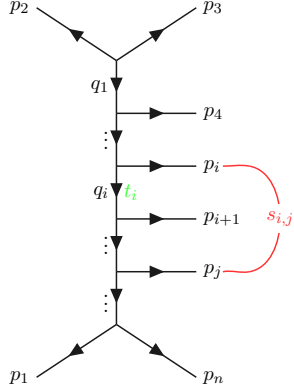}
  \caption{Graphical representation of the choice of Mandelstam variables as defined in Eq.\  (\ref{eq:def_mandelstam}).}
  \label{fig:kinematics}
\end{figure}

In order to describe the multi-Regge limit of a $2\to n-2$
production amplitude, we introduce the standard Mandelstam invariants,
\begin{equation}
\begin{alignedat}{2}
  s_{i,\,j}&=(p_{i}+\cdots+p_{j})^2,\quad && i=3,\dots,n-1, \, j=i+1,\dots,n,\\
  t_i&=(p_2+\cdots+p_{i+2})^2,\quad && i=1,\dots,n-3,
\end{alignedat}
  \label{eq:def_mandelstam}
\end{equation}
where we choose all momenta to be outgoing, as shown in figure \ref{fig:kinematics}.
Note that there are $n-3$ $t$-like
variables $t_i$ and $\frac{1}{2}(n-3)(n-2)$ $s$-like
invariants. The latter include all subenergies, starting from
the variables $s_i = s_{i+2,i+3}$ with $i=1,\dots, n-3$ for pairs of
outgoing particles up to the total energy $s=s_{3,n}$ of the
process. The overall number of these variables is $(n-3) +
\frac{1}{2}(n-3) (n-2) = D_n$. One possible choice of independent
kinematic invariants would include the $2n-6$ variables $t_i$ and $s_i$
along with $n-4$ variables $\eta_a$ which are defined by
\begin{equation}
\eta_{a} := \frac{s_a s_{a+1}}{s_{a+2,a+4}},\quad a = 1, \dots, n-4.
\end{equation}
These are directly related to the subenergies $s_{a+2,a+4}$ for
three outgoing particles. All higher subenergies are then
obtained from the $t_i, s_i$ and $\eta_a$ through the Gram
determinant relations we described above.\par
Because of the dual conformal symmetry of $\mathcal{N}=4$ super Yang-Mills
theory, the remainder function depends on fewer parameters. To
introduce the relevant variables, let us parametrize the momenta $p_i$
through the dual variables $p_i =: x_{i-1} - x_{i}$, with $x_{i+n}=x_i$.
The $x_i$ are position variables for the cusps of a light-like Wilson
loop describing our momentum configuration.
Along with the $x_i$ we also introduce $x_{i,j}^2 = (x_i-x_j)^2$ for
$i,j = 1, \dots, n$, which are just the Mandelstam invariants introduced
before. Note that $x_{i,i}^2 = x_{i,i+1}^2=0$, since we are considering a
light-like Wilson loop. Hence, the $x_{i,j}^2$ provide $D_n = \frac{1}{2}
n(n-3)$ Mandelstam invariants which we may arrange in an $n \times n$
matrix $X_{ij}:=(x_{i,j}^2)$. The $x_{i,j}^2$ possess a rather simple
relation with the Mandelstam invariants we described above. In the case of $n=7$ external gluons, for example, one has
\begin{equation}
X=\begin{bmatrix}
\ 0 \ &\  0 \ &\  x_{1,3}^2\  &\  x_{1,4}^2\  &\  x_{1,5}^2\  &\
x_{1,6}^2\  &\  0\  \\[1mm]
0 & 0 &  0 & x_{2,4}^2 & x_{2,5}^2 & x_{2,6}^2 & x_{2,7}^2 \\[1mm]
x_{1,3}^2  & 0 &  0 & 0 & x_{3,5}^2 & x_{3,6}^2 & x_{3,7}^2 \\[1mm]
x_{1,4}^2  & x_{2,4}^2 &  0 & 0 & 0 & x_{4,6}^2 & x_{4,7}^2 \\[1mm]
x_{1,5}^2  & x_{2,5}^2 &  x_{3,5}^2  & 0 & 0 & 0 & x_{5,7}^2  \\[1mm]
\phantom{mm}  & \ &  \  & \ & 0 & 0 & 0 \\[1mm]
\  & \ &  \  & \ & \ & 0 & 0
\end{bmatrix} \ = \ \begin{bmatrix}
\ 0 \ &\   0 \ &\ \  t_1 \  &\  t_{2}\  &\  t_{3}\  &\
t_{4}\  &\  0\  \\[1mm]
0 & 0 &  0 & \ s_1 \ & \ s_{3,5} \  &\  s_{3,6}\  &\  s\  \\[1mm]
t_{1}  & 0 &  0 & 0 & s_2 & s_{4,6} & s_{4,7} \\[1mm]
t_{2}  & s_1 &  0 & 0 & 0 & s_3 & s_{5,7} \\[1mm]
t_{3}  & s_{3,4} &  s_2  & 0 & 0 & 0 & s_4  \\[1mm]
\phantom{mm}  & \ &  \  & \ & 0 & 0 & 0 \\[1mm]
\  & \ &  \  & \ & \ & 0 & 0
\end{bmatrix},
\label{eq:defX}
\end{equation}
where the entries not shown can be obtained by the symmetry $X_{ij}=X_{ji}$.
For other numbers $n$ of external gluons, the relations take a
similar form, with $t$-like variables in the first row and all the
$s$-like variables filling a triangle in row $2$ to $n-2$.\par
From these Mandelstam invariants we can easily construct $D_n-n$
dual conformal invariant quantities
\begin{equation} \label{eq:CR}
 U_{ij} = \frac{x^2_{i,j+1} x^2_{i+1,j}}{x^2_{i,j}x^2_{i+1,j+1}},
\end{equation}
which are called cross ratios.
Once again, these are not all independent. Additional relations are
obtained from the $c_n = \frac{1}{2}(n-5)(n-6)$ conformal Gram relations
which state that all $7 \times 7$ subdeterminants of the matrix
$X$ must vanish (cf.\ appendix \ref{sec:gram}). They leave us with $D_n
- n - c_n = 3n-15$ independent cross ratios.\par
\begin{figure}[t]
  \centering
  \includegraphics[scale=.7]{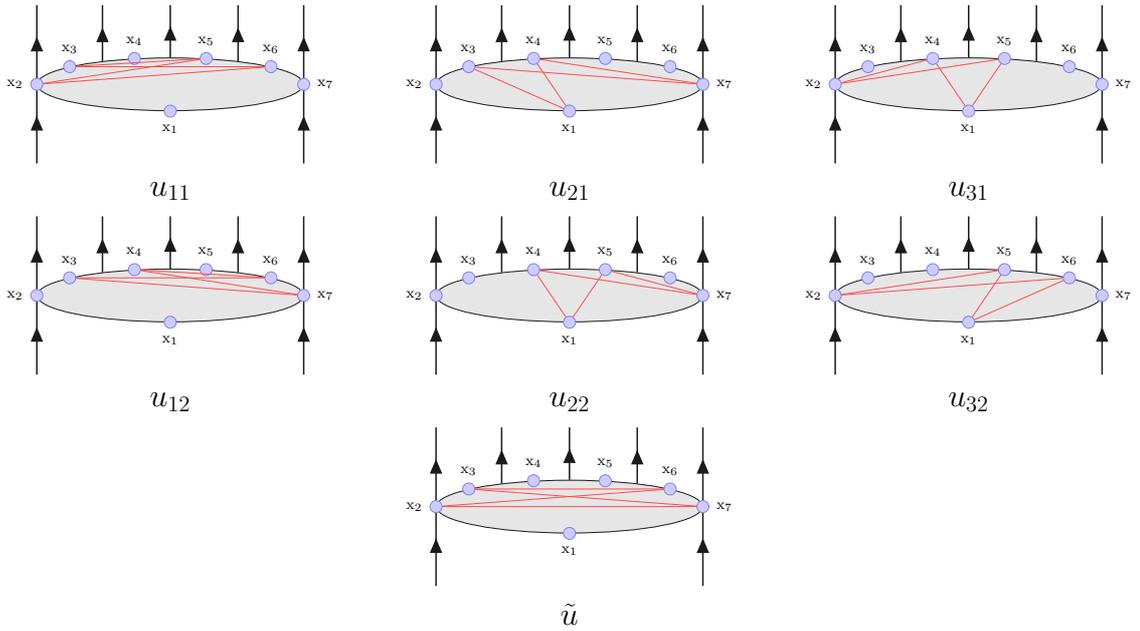}
  \caption{Graphical representation of the cross ratios for the $7$-gluon amplitude. The cross ratios $u_{as}$ are chosen to be independent (cf.\ Eqs.\ (\ref{eq:cr1})-(\ref{eq:cr3})), while $\tilde{u}$ depends on those cross ratios through a conformal Gram relation.}
  \label{fig:7pt_crs}
\end{figure}

As in our discussion of independent Mandelstam invariants, it is
useful to fix an independent set of cross ratios which is adapted
to the multi-Regge limit of a $2 \to n-2$ production
amplitude. In \cite{Bartels:2012gq} we suggest to use
\begin{align}
u_{1\sigma}&= U_{\sigma+1,\sigma+4} =
\frac{x^2_{\sigma+1,\sigma+5}x^2_{\sigma+2,\sigma+4}}
{x^2_{\sigma+2,\sigma+5}x^2_{\sigma+1,\sigma+4}},\label{eq:cr1}\\
u_{2\sigma}&= U_{\sigma+2,n} = \frac{x^2_{\sigma+3,n}x^2_{1,\sigma+2}}
{x^2_{\sigma+2,n}x^2_{1,\sigma+3}},\label{eq:cr2}\\
u_{3\sigma}&= U_{1,\sigma+3} = \frac{x^2_{2,\sigma+3}x^2_{1,\sigma+4}}
{x^2_{2,\sigma+4}x^2_{1,\sigma+3}},
\label{eq:cr3}
\end{align}
where $\sigma=1,\dots,n-5$. All other cross ratios, and in particular
those defined in Eq.\ (\ref{eq:CR}), may be reconstructed from the
$u_{a\sigma}$ by solving the conformal Gram determinant relations (see appendix \ref{sec:gram} for the unique Gram relation in the $7$-gluon case).
A convenient graphical representation of the cross ratios Eqs.\ (\ref{eq:cr1})-(\ref{eq:cr3})
is shown in figure \ref{fig:7pt_crs}.
A symmetry we will use later is target-projectile symmetry, which, as the name suggests, reflects the fact that the amplitude should remain invariant when the two incoming particles $p_1$, $p_2$ are swapped.
In terms of the graphical representation of the cross ratios in figure \ref{fig:7pt_crs}, target-projectile symmetry simply amounts to a reflection on the central vertical axis of each blob. From this, it is easy to deduce that applying target-projectile
symmetry amounts to the following exchange of cross ratios
\begin{equation}
  u_{1\sigma}\ \leftrightarrow \ u_{1(n-4-\sigma)},\quad u_{2\sigma}
  \ \leftrightarrow u_{3(n-4-\sigma)}\ \ .
  \label{eq:target_projectile}
\end{equation}

\subsection{The multi-Regge regime}

By definition, the multi-Regge regime is reached when we impose a strong ordering of the rapidities of the outgoing particles $p_3,\dots,p_n$ while keeping their transverse momenta of the same order. For the Mandelstam variables this translates into all $s$-like
variables becoming large with $t$-like variables fixed. More
precisely, we get a hierarchy
\begin{equation}
  s = s_{3,n} \gg s_{3,n-1}, s_{4,n} \gg \dots \gg s_{i,i+2} \gg s_i \gg -t_i \ .
\end{equation}
In terms of the matrix $X_{ij}$ introduced in Eq.\  (\ref{eq:defX}) this means that the $s$-like variables become larger as we go to the upper right corner.
This scaling of kinematic invariants is discussed in more detail in
\cite{Bartels:2012gq}. In the limiting regime one finds that
\begin {equation} \label{eq:MRLs}
  x_{ij}^2 = s_{i+1,j} \sim s_{i-1} \cdots s_{j-3}\ .
\end{equation}
As we can easily infer from Eq.\  (\ref{eq:MRLs}) and our definition of the independent cross ratios (\ref{eq:cr1})-(\ref{eq:cr3}), the limiting behavior of
the $u_{a\sigma}$ is therefore given by
\begin{equation}
u_{2\sigma} \sim \frac{1}{s_{\sigma+1}}, \quad
u_{3\sigma} \sim \frac{1}{s_{\sigma+1}}, \quad
1-u_{1\sigma} \sim \frac{1}{s_{\sigma+1}}\ .
\label{eq:crs_mrl_behav}
\end{equation}
Hence, the approach to the multi-Regge limit is characterized by the finite, reduced cross ratios
\begin{equation}
  \tilde{u}_{2\sigma}:=\frac{u_{2\sigma}}{1-u_{1\sigma}} =:
\frac{1}{|1+w_\sigma|^2}, \quad
\tilde{u}_{3\sigma}:=\frac{u_{3\sigma}}{1-u_{1\sigma}} =:
 \frac{|w_\sigma|^2}{|1+w_\sigma|^2},
 \label{eq:wparam_lp}
\end{equation}
where we introduced the complex quantities $w_\sigma$ as in \cite{Lipatov:2010ad}.
Let us note that the other cross ratios $U_{ij}$ with $i>1$ and $j<n$
possess the limiting behavior
\begin{equation}
  U_{ij} - 1 \sim s_i^{-1} \cdots s_{j-3}^{-1} \ \ .
\end{equation}
For example, in the case of $n=7$ we can easily see that $\tilde u := U_{26}$ approaches $1$ in the multi-Regge limit, as it must in order to satisfy the
conformal Gram relation we spell out in appendix \ref{sec:gram}.
In fact, inserting $u_{1\sigma}=1$ and $u_{2\sigma}=u_{3\sigma}=0$ into Eq.\  (\ref{eq:gramfull}) leads to the Gram relation
$(\tilde u - 1)^2 = 0$.

\subsection{Multi-Regge regions}
\label{sec:mrl_regions}
As we shall discuss below, it is very important to evaluate the multi-Regge
limit in different kinematical regions. In fact, in the so-called Euclidean region where all kinematic invariants are negative, the
multi-Regge limit of the remainder function vanishes. If this was the only
admissible regime, the multi-Regge limit would be incapable of distinguishing
between the BDS Ansatz and the true scattering amplitude. Fortunately, we
have more freedom in taking the multi-Regge limit.\par
In the Euclidean region, the energy components $p_i^0$ of the four-momenta $p_i$
of the produced particles are taken to be positive. In other
words, all the $p_i$ take values in the forward light cone. But we can admit
different choices for the sign of $p_i^0$. These characterize $2^{n-4}$
admissible kinematical regions\footnote{In this paper, we consider only those regions for which the energy component of the momenta $p_3$ and $p_n$ remains positive. It should be noted that changing the sign of the energy component of those particles is possible as well and considered for example in \cite{Bartels:2013jna}.}. Let us define an associated map
\begin{equation}
\varrho_i =\mathit{sgn}(p_i^0) \in \{\pm 1 \},
\end{equation}
by which we parametrize the region $(\varrho_4 \cdots \varrho_{n-1})$.
It is possible to characterize the regions through the behavior of Mandelstam
invariants and cross ratios. For the Mandelstam invariants, the $t$-like
variables remain negative while $s$-like variables may change sign. Because of Eq.\  (\ref{eq:MRLs}), the
approach to the multi-Regge limit is characterized by the following
behavior
\begin{equation}
  \varrho_{i} \varrho_j \  s_{i,j} \geq 0\,
\end{equation}
with $\varrho_3 = \varrho_{n}=1$.
Furthermore, note that the sign of the $s$-like variables is uniquely
characterized by the signs of the two-particle invariants $s_i$, using
the multi-Regge behavior \eqref{eq:MRLs}.\par
Let us now pass to the cross ratios. In the Euclidean regime, all
cross ratios are positive since they involve an even number of $t$-like
invariants. But once we pass to other regions, the cross ratios
$u_{2\sigma}$ and $u_{3\sigma}$ may change signs according to the
rules
\begin{equation}
\label{eq:uregions}
\varrho_{\sigma+3} \varrho_{\sigma+4}\  u_{2\sigma} \geq 0,
\quad
\varrho_{\sigma+3} \varrho_{\sigma+4}\  u_{3\sigma} \geq 0,
\end{equation}
where $\sigma = 1, \dots, n-5$. Note that the signs of these cross
ratios are not sufficient to specify the regime as they leave one
sign undetermined.\par
We can pass from one multi-Regge region to another by analytic
continuation in the kinematic invariants. In going from the Euclidean
region with all the $u_{a\sigma}$ positive to another region that is
characterized by factors $\varrho_i$, we shall use a curve for which the
cross ratios $U_{ij}$ possess non-vanishing winding number $n_{ij}$
around $U_{ij} = 0$. This winding number is related to the parameters
$\varrho_i$ through\footnote{This formula is simple to derive when using
that the winding number of $s_{i,j}$ is given by $\frac{1}{2}
(1-\varrho_i\varrho_j)$.}
\begin{equation}
\label{eq:winding}
n_{ij} =  \frac{1}{4}(\varrho_{i+1} - \varrho_{i+2})(\varrho_{j}
  - \varrho_{j+1}).
\end{equation}
Winding number $n=1$ means that we encircle the origin of the U-plane in clockwise direction, while $n=-1$ corresponds to a counter-clockwise rotation. The formula for $n_{ij}$ can be used for all $i,j = 1, \dots, n$ if we extend the definition of
$\varrho_i$ such that $\varrho_{1} = \varrho_{2}=\varrho_{n+1}= \varrho_{n+2}=2$.
For the small
cross ratios $u_{2\sigma}= U_{\sigma+2,n}$ and $u_{3\sigma} = U_{1,\sigma+3}$ the
winding number $n_{ij}$ is given by $\pm \frac{1}{2}$. The winding numbers
for all other cross rations are integer valued, i.e.\ these cross ratios
perform full rather than half rotations.\par
The precise curve along which we perform the continuation must be
constructed such that
\begin{itemize}
  \item all winding numbers assume the prescribed values \eqref{eq:winding} and
  \item all conformal Gram determinant relations are respected.
\end{itemize}
For the cross ratios $u_{2\sigma}$ and $u_{3\sigma}$ we propose the following simple
behavior
\begin{equation} \label{eq:rotu23}
  u_{2\sigma}(\varphi) = e^{\frac{i}{2}(\varrho_{\sigma+3} - \varrho_{\sigma+4})
\varphi}\   u_{2\sigma}, \quad
 u_{3\sigma}(\varphi) = e^{\frac{i}{2} (\varrho_{\sigma+4} -
 \varrho_{\sigma+3})\varphi}\
 u_{3\sigma},
\end{equation}
where $\varphi \in [0,\pi]$ parametrizes the curve along which we
continue. This is the simplest behavior that is consistent with our
formula \eqref{eq:winding} and it changes the signs of the cross ratios
$u_{2\sigma}$ and $u_{3\sigma}$ such that we end up in the desired
region, see Eq.\ \eqref{eq:uregions}. For all other cross ratios the
dependence on $\varphi$ is more complicated, at least in general. The
simplest curves that possess the correct winding numbers are, of course,
circles. However, as we shall see in some examples below, having all the
cross ratios move along circles is rarely ever consistent with the Gram
relations. \par
Let us consider some specific examples. In the case of $n=6$ we have four
multi-Regge regions including the Euclidean one. Hence there are three
non-trivial paths connecting the Euclidean region to the three others.
These are given by
\begin{alignat}{4}\label{eq:6pt_pm}
  & P_{6,+-}:\, && u_{11}(\varphi)= u_{11},\quad && u_{21}(\varphi)=e^{i\varphi}u_{21},\quad  && u_{31}(\varphi)=e^{-i\varphi}u_{31}\\[2mm]
  & P_{6,-+}:\, && u_{11}(\varphi)=u_{11},\quad && u_{21}(\varphi)=e^{-i \varphi}u_{21},\quad && u_{31}(\varphi)=e^{i \varphi}u_{31}\\[2mm]
& P_{6,--}:\, && u_{11}(\varphi)=e^{-2i\varphi}u_{11},\quad &&  u_{21}(\varphi)=u_{21},\quad && u_{31}(\varphi)= u_{31},
\label{eq:6pt_mm}\
\end{alignat}
where the subscript $\pm$ indicates the signs of the energies we choose for the
produced particles. Since there are no further cross ratios and no conformal Gram
relations to satisfy, it is obvious that our three curves possess all the
desired properties.\par
If we choose $n=7$ there are eight multi-Regge regions. One is the Euclidean
region, three others correspond to a single change of signs. For the latter,
none of the integer winding numbers $n_{ij}$ with $i>1$ and $j<n$ is actually
non-zero. Such curves will turn out to lead to regions with trivial Regge
limit, both at weak and strong coupling. It therefore suffices to discuss
the remaining four curves. Extending our general prescription in Eq.\
\eqref{eq:rotu23}, the first three of them are given by
\begin{alignat}{5}
& P_{7,+--}:\, && u_{11}(\varphi)=u_{11},\quad && u_{21}(\varphi)=e^{i\varphi}u_{21},\quad && u_{31}(\varphi)=e^{-i\varphi}u_{31}, &&\label{eq:7pt_pmm} \\[1mm]
& && u_{12}(\varphi)=e^{-2i\varphi} u_{12},\quad && u_{22}(\varphi)=u_{22},\quad && u_{32}(\varphi)=u_{32},\quad && \tilde{u}(\varphi)=\tilde{u}\nonumber \\[2mm]
& P_{7,-+-}:\, && u_{11}(\varphi)=e^{2i\varphi} u_{11},\quad && u_{21}(\varphi)=e^{-i \varphi}u_{21},\quad && u_{31}(\varphi)=e^{i \varphi} u_{31}, && \label{eq:7pt_mpm}\\[1mm]
& && u_{12}(\varphi)=e^{2i\varphi}u_{12},\quad && u_{22}(\varphi)=e^{i\varphi} u_{22},\quad && u_{32}(\varphi)=e^{-i\varphi}u_{32},\quad && \tilde{u}(\varphi)=e^{-2i\varphi}\tilde{u} \nonumber \\[2mm]
& P_{7,--+}:\, && u_{11}(\varphi)=e^{-2i\varphi}u_{11},\quad && u_{21}(\varphi)=u_{21},\quad && u_{31}(\varphi)=u_{31}, && \label{eq:7pt_mmp} \\[1mm] \nonumber
& && u_{12}(\varphi)=u_{12},\quad && u_{22}(\varphi)=e^{-i\varphi}u_{22},\quad && u_{32}(\varphi)=e^{i\varphi}u_{32},\quad && \tilde{u}(\varphi)=\tilde{u}\ .
\end{alignat}
It should be noted that the paths spelled out above satisfy the Gram relation only in the multi-Regge limit, i.e.\ when Eq.(\ref{eq:crs_mrl_behav}) holds.
This, however, is satisfied for all cases studied in this paper.
In order to reach the forth region, in which all $\varrho_i = -1$, the conformal
Gram relations force us to consider a more complicated curve. It is still easy to
give an explicit expression in the limit where $u_{2\sigma} = u_{3\sigma} = 0$,
\begin{alignat}{5}
  &P_{7,---}:\, && u_{11}(\varphi)=e^{2i\varphi}\left(1-\sqrt{1-e^{-2i\varphi}}\right)u_{11},\quad && u_{21}(\varphi)=u_{21},\quad && u_{31}(\varphi)=u_{31}, &&\label{eq:7pt_mmm}\\[2mm]
& && u_{12}(\varphi)=e^{2i\varphi}\left(1-\sqrt{1-e^{-2i\varphi}}\right)u_{12},\quad && u_{22}(\varphi)=u_{22},\quad && u_{32}(\varphi)=u_{32},\quad && \tilde{u}(\varphi)=e^{-2i\varphi}\tilde{u}. \nonumber
\end{alignat}
In the multi-Regge limit, the parameters $u_{2\sigma}$, $u_{3\sigma}$ become small so
that $P_{---}$ is a good approximation. This last path illustrates nicely that one
cannot always analytically continue along circles. Just computing the winding numbers
for this path, we find that the cross ratios $u_{a\sigma}$ we choose as independent
variables have winding number zero, while the cross ratio $\tilde{u}$, which is given
in terms of the independent cross ratios via the conformal Gram relation, has winding
number $\tilde n = +1$. This is impossible to realize with $u_{1\sigma}(\varphi) = u_{1\sigma}$
constant because the conformal Gram relations would force $\tilde u$ to be constant, as well.
Choosing the path for $\tilde{u}$ to run along a circle, as we did in our prescription
for $P_{7,---}$, we can solve Eq.\  (\ref{eq:gram_mrl}) for $u_{11}$, $u_{12}$ while
preserving the target-projectile symmetry along the entire path. This leads to the
expressions we have prescribed. Note that the Gram determinant relation (\ref{eq:gram_mrl})
has been derived with the additional assumption that $u_{2\sigma} = u_{3\sigma} = 0$.
Hence, our curve $P_{7,---}$ is only valid in the multi-Regge limit.

\subsection{Weak coupling results}
\label{sec:wc}
On the weak coupling side, soon after the BDS conjecture for the planar $n$-point scattering amplitude
had been published, it became clear that for more than five external legs the BDS formula is incomplete.
The missing pieces are encoded in the remainder functions, $R_n$, and in recent years intense work has
been devoted to the calculation of these functions. The multi-Regge limit provides enormous help in
determining $R_n$.\par
Starting point is the multi-Regge analysis of the $2\to 4$ scattering amplitude in the leading
logarithmic approximation. In \cite{Bartels:2008ce,Bartels:2008sc} it is shown that the failure of the BDS formula is due to the existence of Regge cut contributions. More precisely,
the perturbative analysis shows that the high-energy behavior of the $2\to 4$ scattering
process is described by the exchange of Regge poles - the reggeizing gluon - and
a Regge cut consisting of the exchange of two interacting reggeized gluons. In the planar
approximation this cut is present only in specific kinematic regions which have been named
Mandelstam regions. The BDS formula does not account for this Regge cut contribution.
As shown in \cite{Bartels:2008ce}, it contains the Regge poles, and the phase structure is
correct in the Euclidean region (where all energies are negative) and in the physical region
where all energies are positive. In the Mandelstam region where the Regge cut appears in the perturbative analysis, the BDS formula exhibits a phase which has been identified as the one-loop part of the Regge cut contribution. At the same time, in these
kinematic regions the phase structure of the Regge pole contributions is not correctly
reproduced by the BDS formula \cite{Lipatov:2010qf}.\par
In fact, there is a close connection between the structure of Regge pole contribution and
the existence of Regge cut contributions \cite{Bartels:2013jna}. In the planar approximation,
the Regge pole expression for the $2 \to n-2$ amplitude has a simple factorizing form in only some
kinematic regions, such as the Euclidean region or the region of positive energies. In contrast, just in
those kinematic regions where the Regge cut contributions are present, the Regge pole
expressions develop unphysical singularities which have to be canceled by the Regge cut
singularities. The existence of Regge cuts, therefore, is required to provide a consistent
(i.e. singularity-free and, in the case of $\mathcal{N}=4$ SYM, also conformally invariant) description
of Regge poles.\par
The general analysis of the structure of the Regge pole contributions in $2\to n-2$ scattering
amplitudes is presented in \cite{Bartels:2013jna}. In particular, this analysis allows
to find the analytic form of the Regge pole contributions for planar amplitudes in all kinematic regions,
including the singular pieces which have to be canceled by the Regge cut contributions. Details
are worked out for the cases $2\to4$ and $2\to5$ in \cite{Bartels:2013jna}, and in
\cite{Bartels:yyyy} for the case $2\to 6$.\par
The calculation of the Regge cut contributions is based upon the calculation
of energy discontinuities using unitarity integrals. In order to obtain energy discontinuities
one needs the analytic representation of the scattering amplitudes in multi-Regge kinematics
which, in agreement with the Steinmann relations, exhibits the dependence upon the energy
variables, including the phases in the different kinematic regions. The authors of
\cite{Bartels:xxxx} outline a general strategy for the calculation of Regge cut contributions.
These contain the singular terms which exactly cancel the singular terms of the Regge pole
contributions: when pole and cuts are combined one finds a sum of infrared finite and
conformally invariant pole and cut expressions. Explicit leading order result for the
$2\to 5$ and $2\to6$ cases are obtained in  \cite{Bartels:xxxx} and
\cite{Bartels:yyyy}, respectively.\par
Let us briefly summarize some of those results which are of special interest for the
analysis described in this paper. To be complete we begin with the $2\to 4$ case.
The Regge cut appears in the kinematic region $(--)$: this is the one which is
studied in \cite{Bartels:2010ej} and it is contained in the list (\ref{eq:6pt_pm}) -
(\ref{eq:6pt_mm}). In terms of Mandelstam variables the energies have the following
sign structure
\begin{equation}
s,s_2>0,\,s_1,s_3, s_{3,5},s_{4,6}<0\ .
\end{equation}
Note that there is a second region, not mentioned in the above list, in which the cut
appears: it is obtained by `twisting' the $t_2$ channel. The other two regions, $(+-)$
and $(-+)$, do not contain the Regge cut contribution. The corresponding path of analytic
continuation for the region $(--)$ is described in Eq.\ (\ref{eq:6pt_mm}). It is possible
to define other, more complicated paths which connect the same starting and final values
of energies, but the path Eq.\ (\ref{eq:6pt_mm}) appears to be the simplest one. The Regge cut
amplitude which emerges after combination with the Regge pole contribution is
derived in \cite{Bartels:2008sc} and the full remainder function in this region
is obtained in \cite{Bartels:xxxx}:
\begin{equation}
\left[ e^{R_6 + i\delta_{6,--}}\right]_{--} = \cos \pi \omega_{ab} + i\delta_{6,--} +2i f_{\omega_2}\ .
\end{equation}
In terms of the cross ratios the cut amplitude $f_{\omega_2}$ has the form:
\begin{equation}
\label{short2-cut}
f_{\omega_2}=
\frac{g^2 N_c}{16\pi^2} \sum_n (-1)^n  \left( \frac{w}{w^*}\right)^{\frac{n}{2}}
\int \frac{d\nu}{2\pi i} \Phi_{\nu,n}^*
\left[ \left(-\sqrt{u_2u_3}\right)^ {-\omega(\nu,n)} - 1 \right] \Phi_{\nu,n}
|w|^{2i\nu}  ,
\end{equation}
where $w=w_1$ was introduced in Eq.\ (\ref{eq:wparam_lp}).
Here $\Phi_{\nu,n}$ denotes the impact factor and $\omega(\nu,n)$ is the eigenvalue function of the color octet BFKL Hamiltonian.
These quantities are known up to next-to-leading order from direct field theory calculations \cite{Bartels:2008sc, Lipatov:2010ad, Fadin:2011we}.
Using the bootstrap program of \cite{Dixon:2013eka, Dixon:2014voa}, the BFKL eigenvalue and the impact factor can be determined up to NNLLA and N$^3$LLA, respectively, and all-order expressions derived from the Wilson loop OPE are put forward in \cite{Basso:2014jfa}.\par
In \cite{Lipatov:2009nt} it is shown that the BFKL Hamiltonian is equivalent to the
non-compact $SL(2,C)$ Heisenberg Hamiltonian for an open spin chain, with $\omega(\nu,n)$ being the eigenvalue function for the spin chain of length two.\par
Next we turn to the case $2\to 5$ \cite{Bartels:xxxx}.  Here, the weak-coupling analysis leads
to three different Regge cut contributions composed of two reggeized gluons, two `short' cuts
in the $t_2$ and $t_3$ channels, respectively, and one `long cut' extending over the $t_2$
and the $t_3$ channels. The different kinematic regions in which these cuts appear are listed
in Eqs.\ (\ref{eq:7pt_pmm})-(\ref{eq:7pt_mmm}), together with simple choices of paths of analytic
continuation. The simplest cases are $(+--)$ and $(--+)$: they contain only the short cuts in
the $t_3$ and the $t_2$ channels, respectively. In these regions the remainder function has the
following form
\begin{align}
\left[ e^{R_7 + i\delta_{7,+--}}\right]_{+--}&=\cos\pi\omega_{bc}+i \delta_{7,+--}+ 2if_{\omega_3}\\
\left[ e^{R_7 + i\delta_{7,--+}}\right]_{--+}&=\cos\pi\omega_{ab}+i \delta_{7,--+} +2if_{\omega_2}
\end{align}
The integral expressions for the Regge cut amplitudes $f_{\omega_2}$ and $f_{\omega_3}$
are easily obtained from Eq.\ (\ref{short2-cut}) by substituting the cross ratios
$u_i\to u_{i\sigma}$. For the remaining two regions also the long cut contributes.
For $(---)$ and $(-+-)$ the remainder functions are given by
\begin{align}
\left[e^{R_7 + i\delta_{7,---}}\right]_{---}&= \cos\pi\omega_{ac}+i\delta_{7,---}+ 2i  f_{\omega_2\omega_3}\\
\left[e^{R_7 + i\delta_{7,-+-}}\right]_{-+-} &= e^{i\pi \omega_{ba}} e^{i\pi \omega_{bc}} +
i \delta_{7,-+-}  +2i(f_{\omega_2\omega_3}
-e^{i\pi \omega_{bc}}f_{\omega_2}-e^{i\pi \omega_{ba}}f_{\omega_3}).
\end{align}
 The integral for the long cut reads
\begin{eqnarray}
\label{long2-cut}
f_{\omega_2\omega_3}= &&\frac{a}{2} \sum_{n_1,n_2} (-1)^{n_1+n_2}
\left(\frac{w_1}{w_1^*}\right)^{n_1} \left(\frac{w_2}{w_2^*}\right)^{n_2}
\int \frac{d\nu_1 d\nu_2}{(2\pi)^2}
 \Phi(\nu_1,n_1)^*  |w_1|^{2i\nu_1}
\left(-\sqrt{u_{21} u_{31}}
 \right)
^{-\omega(\nu_1,n_1)}
\nonumber\\[2mm]
&&C(\nu_1,\nu_2,n_1,n_2) \left(-\sqrt{u_{22} u_{32}}
 \right)
^{-\omega(\nu_2,n_2)}
|w_2|^{2i\nu_2} \Phi(\nu_2,n_2) |_{\text{sub}}.
\end{eqnarray}
Here the impact factor $\Phi(\nu,n)$ is the same as in Eq.\ (\ref{short2-cut}), and
$C(\nu_1,\nu_2,n_1,n_2)$ stands for the central production vertex.  The latter is known
in leading order \cite{Bartels:2011ge}. The subscript $|_{sub}$ indicates that we have
subtracted the one-loop contribution.  In \cite{Bartels:xxxx} a few more kinematic regions
with Regge cut contributions are listed; they will not be mentioned here.\par
It is important to note that in leading order the integral representation is real-valued up to the phase factors $e^{-i\pi\omega(\nu_i,n_i)}$.
Beyond leading order this will no longer be the case. As explained in \cite{Bartels:xxxx},
one can still write the Regge cut amplitude in the form of Eq.\ (\ref{long2-cut}), but with a
complex-valued expression for the production vertex $C$. Alternatively, the amplitude
breaks into two pieces with different phase factors.\par
Before we turn to the strong coupling regime let us make a final remark on the weak coupling
results summarized in this section. Comparing the results for the $2\to4$ scattering amplitude with
those of the $2\to5$ case we would like to stress the close connections. Obviously the
short Regge cuts in the $2\to5$ case have the same functional form as the $2\to4$
cut contribution (with suitable replacements of the cross ratios), i.e. the same impact factor
and eigenvalue function $\omega(\nu,n)$. Also the expression for the long cut
contains - apart from the new production vertex - the same building blocks.\par
New elements appear in the $2 \to 6$ scattering amplitude: in the region $(-++-)$ one finds a Regge cut consisting of three reggeized gluons. Apart from a new impact factor, a new eigenvalue function $\omega_3(\nu_1,n_1;\nu_2,n_2)$ appears which belongs to a spin chain of length three and can be parametrized by two sets of conformal quantum numbers. In leading order this eigenvalue function can be written as a sum of the two functions $\omega(\nu_1,n_1)$ and $\omega(\nu_2,n_2)$ \cite{Lipatov:2009nt,Derkachov:2014gya}. Whether this simple additivity remains valid also beyond leading order is presently not known.

\section{Scattering amplitude at strong coupling}
\label{sec:gen_alg}
In this section we shall outline the general algorithm for the calculation
of the remainder function of strongly coupled $\mathcal{N}=4$ SYM in the
multi-Regge limit. All the key elements of our description are applicable
for any number $n$ of external gluons. There are a few formulas that
we shall only spell out for $n=7$ because their precise form for other
values of $n$ is irrelevant both for the presentation of the general
algorithm and for the example we shall work out in the next section.
After reviewing the form of the scattering amplitude and its relation
with a special set of thermodynamic Bethe Ansatz equations, we will
discuss how to perform the multi-Regge limit. In particular we shall
show that the multi-Regge limit is a particular large mass limit in
the underlying thermodynamic Bethe Ansatz. In such a large mass limit,
the thermodynamic Bethe Ansatz equations get replaced by a much simpler
set of algebraic Bethe Ansatz equations. The latter will be described
explicitly in the third subsection. We conclude the section by illustrating
the general algorithm through the simplest example, namely the case of
$n=6$ gluons, and explain how to evaluate the various contributions to the remainder
function in the multi-Regge limit. For the least trivial contribution
to the remainder function this will lead to a set of Bethe Ansatz equations. Special
solutions of these Bethe Ansatz equations are associated with the various
kinematical regions discussed above.

\subsection{Amplitude and thermodynamic Bethe Ansatz}
\label{sec:rv_sc_amps}
The problem of calculating strong coupling scattering amplitudes to leading order was solved in a series of papers \cite{Alday:2007hr,Alday:2009yn, Alday:2009dv,Alday:2010vh}, the solution of which we will briefly review here.

It turns out that the leading order of the $n$-gluon scattering amplitude can
be written as
\begin{equation}\label{amplitude}
  \mathcal{A}\ \sim \ e^{-\frac{\sqrt{\lambda}}{2\pi}\mathrm{A}}
  = e^{-\frac{\sqrt{\lambda}}{2\pi} \mathrm{A}_\text{BDS} + R}
\end{equation}
where the quantity $\mathrm{A}$ consists of several different terms\footnote{For the special case $n=4k$ there is an additional contribution $\mathrm{A}_\mathrm{extra}$ which is worked out in detail in \cite{Yang:2010az}. Since this case is not considered in this paper, we drop this contribution in Eq.(\ref{amplitude}).} to be discussed
one by one,
\begin{equation} \label{eq:A4term}
  \mathrm{A}=\mathrm{A}_{\mathrm{div}}+\mathrm{A}_{\mathrm{free}}+
  \mathrm{A}_{\mathrm{BDS-like}}+\mathrm{A}_{\mathrm{per}}\ .
\end{equation}
Let us discuss the four different contributions in the order of their
appearance. To begin with, $\mathrm{A}_{\mathrm{div}}$ encodes the infrared
divergences of the amplitude and it reads
\begin{equation}
  \mathrm{A}_{\mathrm{div}}\ =\ \frac{1}{8}\sum\limits_i\log^2
  \left( \varepsilon^2 x^2_{i,i+2} \right),
  \label{eq:adiv}
\end{equation}
where $\varepsilon$ is a radial cutoff for AdS$_5$. The next term
$\mathrm{A}_\mathrm{free}$ is the most interesting and least trivial
one. As we anticipated in the introduction, it is constructed from a
one-dimensional auxiliary quantum integrable system. For the $n$-gluon
amplitude, it is defined in terms of $3n-15$ functions, $\Yf_{a,s}(\theta)$,
where $a=1,2,3$ and $s=1,\dots,n-5$. These functions are determined by a
set of non-linear integral equations, called the Y-system equations,
\begin{align}
\mathrm{log}\Yf_{1,s}&=-m_s\cosh\theta-C_s-\frac{1}{2}K_2\star\beta_s- K_1\star\alpha_s-\frac{1}{2}K_3\star\gamma_s,\label{eq:ysys1}\\[2mm]
\mathrm{log}\Yf_{2,s}&=-m_s\sqrt{2}\cosh\theta-K_2\star\alpha_s-K_1\star\beta_s,\label{eq:ysys2}\\[2mm]
\mathrm{log}\Yf_{3,s}&=-m_s\cosh\theta+C_s-\frac{1}{2}K_2\star\beta_s- K_1\star\alpha_s+\frac{1}{2}K_3\star\gamma_s,\label{eq:ysys3}
\end{align}
where $m_s$ and $C_s$ are parameters that one can think of as mass parameters and
chemical potentials of excitations in the one-dimensional auxiliary quantum system.
The objects $\alpha_s$, $\beta_s$, $\gamma_s$ that appear on the right hand side
are defined in terms of the functions $\Yf_{a,s}$ as
\begin{align}
\alpha_s&=\mathrm{log}\frac{(1+\Yf_{1,s})(1+\Yf_{3,s})}{(1+\Yf_{2,s-1})(1+\Yf_{2,s+1})},\label{eq:alpha}\\[2mm]
\beta_s&=\mathrm{log}\frac{(1+\Yf_{2,s})^2}{(1+\Yf_{1,s-1})(1+\Yf_{1,s+1})(1+\Yf_{3,s-1})(1+\Yf_{3,s+1})},\label{eq:beta}\\[2mm]
\gamma_s&=\mathrm{log}\frac{(1+\Yf_{1,s-1})(1+\Yf_{3,s+1})}{(1+\Yf_{1,s+1})(1+\Yf_{3,s-1})}. \label{eq:gammaY}
\end{align}
In the Y-system these combinations of $\Yf$-functions are convoluted with the
following simple kernel functions
\begin{align} \label{eq:kernels}
K_1=\frac{1}{2\pi}\frac{1}{\cosh\theta}, \quad
K_2=\frac{\sqrt{2}}{\pi}\frac{\cosh\theta}{\cosh 2\theta},
\quad K_3=\frac{i}{\pi}\tanh 2\theta.
\end{align}
These kernel functions can be thought of as describing the (integrable)
interaction between the excitations of the auxiliary quantum system. Let
us also recall that the convolution product is defined as
\begin{align}
(K \ast f)(\theta) \ =  \ \int\limits_{-\infty}^{\infty} d\theta'
K(\theta-\theta')f(\theta')\ .
\end{align}
The spectral parameter $\theta$ in the Y-system equations is a complex
variable. From the definition \eqref{eq:kernels} of the kernel functions
it is clear that for certain values of $\theta$, the kernels can become
singular. Therefore, the form of the Y-system as presented in Eqs.\ (\ref{eq:ysys1})-(\ref{eq:ysys3}) is only valid for $\left|\mathrm{Im}\,
\theta\right|\leq\frac{\pi}{4}$. If we wish to evaluate the Y-system for
larger imaginary parts of $\theta$, we have to pick up residues or use a
powerful recursion relation for the Y-functions,
\begin{equation} \label{eq:recursion}
\Y{a}{s}{r}=\frac{\left(1+\Y{a}{s+1}{r+1}\right)
\left(1+\Y{4-a}{s-1}{r+1}\right)}{\Y{4-a}{s}{r+2}
\left(1+\frac{1}{\Y{a+1}{s}{r+1}}\right)\left(1+\frac{1}{\Y{a-1}{s}{r+1}}\right)},
\end{equation}
where $\Y{a}{s}{r}(\theta) = \Yf_{a,s}(\theta + i r \pi/4)$ denotes a shift in
$\theta$ by a multiple of $i\frac{\pi}{4}$. When using the recursion relation,
it should be noted that $\Yf_{0,s}=\Yf_{4,s}=\infty$, as well as $\Yf_{a,0}=\Yf_{a,n-4}=0$.\\
An additional modification of the Y-system equations has to be made when the
parameters $m_s$ become complex, i.e. $m_s=|m_s|e^{i\phi_s}$. In this case,
we have to substitute
\begin{align}
  & m_s \rightarrow\left|m_s\right|,\quad \Yf_{a,s}(\theta)\rightarrow \tilde{\Yf}_{a,s}(\theta):=\Yf_{a,s}(\theta+i\phi_s),\\[2mm]   & K^{a,a'}_{s,s'}(\theta-\theta')\rightarrow K^{a,a'}_{s,s'}(\theta-\theta'+i(\phi_s-\phi_{s'})),
\end{align}
in Eqs.\ (\ref{eq:ysys1})-(\ref{eq:ysys3}), as is shown in
\cite{Alday:2010vh}. Once we solve the Y-system equations for the
$\tilde{\Yf}_{a,s}$, we can finally calculate
$\mathrm{A}_{\mathrm{free}}$, which reads
\begin{equation}
  \mathrm{A}_{\mathrm{free}}=\sum\limits_s
  \int\frac{d\theta}{2\pi}|m_s|\cosh\theta\left[\left(1+\tilde{\Yf}_{1,s} \right)\left(1+\tilde{\Yf}_{3,s}\right)
  \left(1+\tilde{\Yf}_{2,s}\right)^{\sqrt{2}}\right](\theta).
\label{eq:afree}
\end{equation}
The expression resembles similar formulas for the free energy in
one-dimensional integrable quantum systems. As is stands, the expression
for $\mathrm{A}_\mathrm{free}$ provides a function of the system
parameters $m_s$, $C_s$ and $\phi_s$. We shall review below how these
are related to the physical cross ratios.

There are two additional terms in the general expression \eqref{eq:A4term}
for the logarithm of the scattering amplitude, namely $\mathrm{A}_\mathrm{BDS-like}$
and $\mathrm{A}_\mathrm{per}$. These are again much simpler to spell out than $\mathrm{A}_\mathrm{free}$ and
although they are known for any number $n$ of gluons, we shall content
ourselves with the expressions for $n=7$. In this case,
$\mathrm{A}_{\mathrm{BDS-like}}$ can be written as
\begin{equation}
  \mathrm{A}_{\mathrm{BDS-like}}=-\frac{1}{4}\sum\limits_{i=1}^{7}\left( \log^2 x^2_{i,i+2}+ \sum\limits_{k=0}^{2}(-1)^{k+1}\log x^2_{i,i+2}\log x^2_{i+2k+1,i+2k+3}\right).
  \label{eq:abdslike}
\end{equation}
It is customary to subtract the one-loop finite part of the BDS amplitude $\mathrm{A}_{\mathrm{BDS}}$ that was written down in \cite{Bern:2005iz}.
Since both $\mathrm{A}_{\mathrm{BDS}}$ and $\mathrm{A}_{\mathrm{BDS-like}}$
satisfy the same anomalous Ward identity of dual conformal invariance \cite{Drummond:2007au}, their
difference can only be a function of the conformal cross ratios introduced
in section \ref{sec:kinematics}. For seven points this is written down in \cite{Yang:2010as} and reads
\begin{align}
  \nonumber\Delta:=&\mathrm{A}_{\mathrm{BDS-like}}-\mathrm{A}_{\mathrm{BDS}}
  =-\frac{1}{4}\sum\limits_{i=1}^{7}\left(\log^2 u_i+\mathrm{Li}_2(1-u_i)\right)+\frac{1}{8}\log u_{11}\log\left(\frac{u_{21}u_{22}}{\tilde{u}\,u_{32}}\right)\\
   &+\frac{1}{8}\log u_{12}\log\left(\frac{u_{32}u_{31}}{\tilde{u}\,u_{21}}\right)+\frac{1}{8}\log u_{21}\log\left(\frac{u_{11}u_{32}}{u_{12}u_{22}}\right)+\frac{1}{8}\log u_{22}\log\left(\frac{u_{11}\,\tilde{u}}{u_{21}u_{31}}\right)\label{eq:deltabc}\\
&+\frac{1}{8}\log u_{31}\log\left(\frac{u_{12}\,\tilde{u}}{u_{22}u_{32}}\right)+\frac{1}{8}\log u_{32}\log\left(\frac{u_{12}u_{21}}{u_{11}u_{31}}\right)+\frac{1}{8}\log \tilde{u}\log\left(\frac{u_{22}u_{31}}{u_{11}u_{12}}\right).\nonumber
\end{align}
The last missing piece of the amplitude is $\mathrm{A}_{\mathrm{per}}$,
which is a function of the auxiliary parameters $m_s$ and $\phi_s$ and
is given by
\begin{equation}
  \mathrm{A}_{\mathrm{per}}=\frac{|m_1|^2}{2}+\frac{|m_2|^2}{2}+
  \frac{1}{\sqrt{2}}|m_1||m_2|\left(\cos\phi_1\cos\phi_2+\sin\phi_1\sin\phi_2\right)
  \label{eq:aper}
\end{equation}
for the $7$-point amplitude. Once again, $\mathrm{A}_\mathrm{per}$ becomes a
function of the cross ratios once we understand how the system parameters are
related to the kinematics, see next subsection. Summing things up, we can now
write the remainder function $R$ that was defined in Eq.\ \eqref{amplitude} as
\begin{equation}
  e^R:=e^{-\frac{\sqrt{\lambda}}{2\pi}\left(\Delta+\mathrm{A}_{\mathrm{free}}
  +\mathrm{A}_{\mathrm{per}}\right)}\ .
  \label{eq:def_remainder_fct}
\end{equation}
This concludes our review of the remainder function in strongly
coupled $\mathcal{N}=4$ SYM theory. We can now begin to approach the multi-Regge
regime.

\subsection{Approaching multi-Regge kinematics}
\label{sec:ysys_mrl}

As we have stressed in the previous subsection the two terms
$\mathrm{A}_{\mathrm{free}}$ and $\mathrm{A}_{\mathrm{per}}$ are still
written as functions of the system parameters $m_s, C_s$ and $\phi_s$
rather than the cross ratios $u_{a\sigma}$. Notice that the number $3n-15$
of system parameters coincides with the number of independent cross ratios.
Indeed, the two sets of variables can be mapped onto each other. As explained
in \cite{Alday:2010vh}, the cross ratios are connected to Y-functions
evaluated at special values of the spectral parameter $\theta$. More
precisely, we have that
\begin{equation}
  U_{ij}=\U{i-j-2}{i+j+2},
  \label{eq:rel_crs_yfcts}
\end{equation}
where $Y_{a,s}^{[k]}$ is shorthand for $\Y{a}{s}{k}(0)=
\mathrm{Y}_{a,s}\left(ik\frac{\pi}{4}\right)$.
Since the right-hand side depends only on the Y-system parameters,
relation (\ref{eq:rel_crs_yfcts}) can be inverted to determine
their dependence on the cross ratios, i.e.\ on the kinematics of our
scattering problem.\par
Imposing multi-Regge behavior for the cross ratios, we are driven to
very special values of the Y-system parameters. Indeed, in
\cite{Bartels:2012gq} we find that the following limiting behavior
of the system parameters
\begin{equation}
  |m_s|\rightarrow\infty,\quad \phi_s\rightarrow (1-s)\frac{\pi}{4},\quad C_s=\mathrm{const.}\in i\mathbb{R}
  \label{eq:ysys_params_mrl}
\end{equation}
leads to multi-Regge behavior as given in Eqs.\ (\ref{eq:crs_mrl_behav})
for the cross ratios. More specifically, introducing the parameters
\begin{equation}
  \varepsilon_s=e^{-|m_s|\cos\left(\phi_s-(1-s)\frac{\pi}{4}\right)},\quad w_s=e^{|m_s|\sin\left(\phi_s-(1-s)\frac{\pi}{4}\right)}
  \label{eq:params_eps_w}
\end{equation}
the cross ratios behave as
\begin{align}
  u_{1\sigma}&=1-\varepsilon_{n-4-\sigma}\left(w_{n-4-\sigma}+\frac{1}{w_{n-4-\sigma}}+2\cosh C_{n-4-\sigma}\right),\label{eq:behaviour_crs_params1}
\\[2mm]
  u_{2\sigma}&=\varepsilon_{n-4-\sigma}w_{n-4-\sigma},\\[2mm]
  u_{3\sigma}&=\frac{\varepsilon_{n-4-\sigma}}{w_{n-4-\sigma}},
  \label{eq:behaviour_crs_params3}
\end{align}
where the $\varepsilon_s$ go to zero in the multi-Regge limit, while the
$w_s$ attain a constant value.
These equations are valid up to higher corrections in the $\varepsilon_s$.
The reason for this simplification is, again as shown in \cite{Bartels:2012gq},
that in the multi-Regge limit the integrals in Eqs.\ (\ref{eq:ysys1})-(\ref{eq:ysys3})
can be neglected, although one possibly needs to pick up residue contributions,
depending on the value of $\theta$.\par
This is already enough information to calculate the remainder function in the multi-Regge
regime before any cross ratio is analytically continued, i.e.\ in the Euclidean region.
We know from \cite{Bartels:2008ce,Bartels:2008sc} that the remainder function must be trivial
in this multi-Regge region and we show in appendix \ref{sec:remeuc} that it is
indeed the case. The triviality of the remainder function is closely related to that
of the free energy. The latter has a beautiful physical interpretation. In the large
mass limit, the quantum fluctuations in the auxiliary one-dimensional quantum system
are suppressed. Hence the vacuum becomes trivial and so does the free energy. In the
next subsection we shall explain how the one-dimensional quantum system manages to
produce less trivial results for the other regions.
\par
Before going there, let us briefly spell out the most important formulas of
this subsection in the case of $n=7$ since we need these expressions for our
analysis in the final section. In this case, we have six Y-functions which
determine the six independent cross ratios through
\begin{equation}
\begin{aligned}
  u_{11}&=\U{2}{2},&\, u_{21}&=\U{2}{-2},&\, u_{31}&=\U{2}{0},\\
u_{12}&=\U{1}{-3},&\, u_{22}&=\U{1}{-1},&\, u_{32}&=\U{1}{1}.
\end{aligned}
\label{eq:crs}
\end{equation}
The values of the Y-system parameters in the multi-Regge limit are given
by
\begin{equation}
|m_s|\rightarrow\mathrm{ large},\quad \phi_1\rightarrow 0,\quad \phi_2\rightarrow -i\frac{\pi}{4},\quad C_s=\mathrm{ const.},
\label{eq:limit_mrl}
\end{equation}
and from Eqs.\ \eqref{eq:behaviour_crs_params1}-\eqref{eq:behaviour_crs_params3} we find the behavior of the cross
ratios to take the form
\begin{equation}
\begin{aligned}
u_{11}&=1-\left(w_2+\frac{1}{w_2}+2\cosh C_2\right)\varepsilon_2,&\, u_{21}&=w_2\varepsilon_2,&\, u_{31}&=\frac{\varepsilon_2}{w_2},\\
u_{12}&=1-\left(w_1+\frac{1}{w_1}+2\cosh C_1\right)\varepsilon_1,&\, u_{22}&=w_1\varepsilon_1,&\, u_{32}&=\frac{\varepsilon_1}{w_1}.
\end{aligned}
\label{eq:crs7pt}
\end{equation}
This is all the input we need for the computation of the remainder function in
various Regge regions in section \ref{sec:calcamp}.

\subsection{Amplitude and multi-Regge Bethe Ansatz}
\label{sec:ba}

As mentioned in the previous subsection, the non-linear integral equations
that control the $n$-gluon amplitude at strong coupling simplify drastically
when we take the multi-Regge limit. In fact, in the limiting regime we can
actually neglect the integral contributions. Such a limit is well known in
the theory of integrable systems. It corresponds to an infrared limit in
which the solution of the integrable model boils down to solving a set
of algebraic Bethe Ansatz equations.\par
Before performing the multi-Regge limit, the Y-system for scattering
amplitudes of strongly coupled $\mathcal{N}=4$ SYM theory takes the
general form
\begin{equation}
  \log \tilde \Yf_{a,s} (\theta) =  - p_{a,s}(\theta) + \sum_{a',s'}
  \int d\theta' K_{s,s'}^{a,a'}(\theta-\theta'+i\phi_s-i\phi_{s'})
   \log \left(1 + \tilde \Yf_{a',s'}(\theta')\right).
\label{TBA}
\end{equation}
The source terms $p_{a,s}$ and the kernels $K_{s,s'}^{a,a'}$ depend
on the $3n-15$ parameters $|m_s|, C_s$ and $\phi_s$ in a way that was spelled
out in the previous subsection. The expressions for the source terms are easy
to state,
\begin{equation}
  p_{2,s}(\theta) = \sqrt{2}|m_s|\cosh\theta, \quad
p_{2\pm 1,s} (\theta) = |m_s| \cosh\theta \mp C_s\
\end{equation}
and we leave it to the reader to infer formulas for the kernels from
our discussion above. As we explained before, we can use Eqs.\ \eqref{TBA}
as long as $|\phi_s-\phi_{s+1}| < \pi/4$. Once we have solved for the
$\tilde \Yf_{a,s}$ we can compute the free energy through the formula
Eq.\ (\ref{eq:afree}). In all these formulas, integrations are performed
over the real axis. \par
We are now prepared to review how Bethe Ansatz equations emerge from
the Y-system. In order to do so, we
represent the kernel functions $K_{s,s'}^{a,a'}$ through new objects
$S_{s,s'}^{a,a'}$,
\begin{equation}
  -2\pi i K_{s,s'}^{a,a'}(\theta) =: \partial_\theta \log S_{s,s'}^{a,a'}(\theta).
\label{eq:def_smat}  \end{equation}
As specific examples, the S-matrices corresponding to the basic kernels Eq.\ (\ref{eq:kernels}) take the form
\begin{equation}
  S_1(\theta)=i\frac{1-ie^\theta}{1+ie^\theta},\quad S_2(\theta)=\frac{2i\sinh\theta-\sqrt{2}}{2i\sinh\theta+\sqrt{2}},\quad S_3(\theta)=\cosh2\theta.
  \label{eq:basic_smat}
\end{equation}
As Dorey and Tateo \cite{Dorey:1996re,Dorey:1997rb} observed in the context of
one-dimensional integrable systems, the form of the Y-system equations can
change as one deforms the system parameters. In fact, upon analytic continuation
of the parameters $|m_s|$, $C_s$ and $\phi_s$ of the Y-system some of the
solutions to the equations $\tilde \Yf_{a,s}(\theta) = -1$ may cross the
real axis. We shall enumerate those solutions by an index $\nu$
\begin{equation} \label{poleeq}
 \tilde \Yf_{a,s}(\theta^{(a,s)}_\nu) = -1,\mathrm{for}\ \nu = 1,\dots, N_{a,s}
  \ \ .
\end{equation}
When this happens, the integral on the right-hand side of Eq.\
\eqref{TBA} picks up a residue term since there is a pole crossing
the integration contour. Hence, after analytic continuation the
equations \eqref{TBA} take the form
\begin{align}
\log \tilde \Yf'_{a,s}(\theta)=&-p'_{a,s}(\theta) +
\sum_{a',s'} \sum_{\nu = 1}^{N_{a',s'}}{\mu^{(a',s')}_\nu}\log S_{s,s'}^{a,a'}(\theta-\theta^{(a',s')}_\nu+i\phi'_s-i\phi'_{s'})\\
&+\sum_{a',s'} \int d\theta' K_{s,s'}^{a,a'}(\theta-\theta'+i\phi'_s-i\phi'_{s'})\log \left(1 + \tilde \Yf'_{a',s'}(\theta')\right).
\label{TBAex}
\end{align}
Here, the sign factors $\mu^{(a,s)}_\nu \in \{ \pm 1\}$ depend on
whether the solution of Eq.\  \eqref{poleeq} crosses from the lower
into the upper half-plane or in the opposite direction. At the
endpoint of the continuation, the system parameters assume
the values $|m_s|'$, $C'_s$ and $\phi'_s$, which may differ from those
we started with. Therefore, we place a prime on all quantities
that are defined in terms of the system parameters. Later we
shall impose the condition that our continuation corresponds
to a specific curve in the space of cross ratios (cf.\ section
\ref{sec:mrl_regions}). That allows us to determine the primed
system parameters. Before we do so, let us now send the system
parameters of our non-linear integral equations into a regime
where the integrals can be neglected, e.g.\ into the multi-Regge
regime, where the Eqs.\  \eqref{TBAex} become
\begin{equation}
\log \tilde \Yf'_{a,s}(\theta) = - p'_{a,s}(\theta) +
\sum_{a',s'} \sum\limits_{\nu = 1}^{N_{a',s'}} {\mu^{(a',s')}_\nu}
\log S_{s,s'}^{a,a'}(\theta-\theta^{(a',s')}_\nu + i\phi'_s-i\phi'_{s'}).
\label{TBAlogBA}
\end{equation}
We can exponentiate this set of equations for the functions
$\tilde \Yf'_{a,s}(\theta)$ and insert the values $\theta =
\theta^{(a,s)}_\nu$ satisfying Eq.\ \eqref{poleeq}
to obtain
\begin{equation}
-e^{ p'_{a,s}(\theta^{(a,s)}_\mu)}=\prod_{a',s'} \prod_{\nu = 1}^{N_{a',s'}}S_{s,s'}^{a,a'}(\theta^{(a,s)}_\mu-
\theta^{(a',s')}_\nu+i\phi'_s-i\phi'_{s'})^{\mu^{(a',s')}_\nu}.
\label{TBABA}
\end{equation}
In our context, these equations simply determine the possible location
of the solutions $\theta^{(a,s)}_\nu$ to Eqs.\ (\ref{poleeq}) and we
will call them \textit{endpoint conditions} in the following. The form
of the equations coincides with a usual Bethe Ansatz which imposes
single-valuedness of wave functions for a dilute system of particles
on a one-dimensional circle of radius $R$. The term $e^{p'_{a,s}}$
accounts for the phase shift of a freely moving particle with momentum $ k'_{a,s}(\theta^{(a,s)}_\nu)= -i p'_{a,s}(\theta^{(a,s)}_\nu)/R$ when
we take it once around a circle of radius $R$. The remaining factors
arise from the scattering with other particles that may be distributed
along the one-dimensional circle. Hence, the quantities $S_{s,s'}^{a,a'}$
introduced in Eq.\ \eqref{eq:def_smat} are interpreted as scattering
matrices for excitations of some integrable system and the source terms
$k'_{a,s}$ describe the momentum.\par
Once we have found a solution for the Eqs.\  \eqref{TBABA}, we can
insert it back into the Eqs.\ \eqref{TBAlogBA} to determine the
functions $\tilde \Yf$. From these $\tilde \Yf$-functions we then
compute the cross ratios. This may require repeated use of the
recursion relations \eqref{eq:recursion} for the $\Yf$-functions,
but it is straightforward. The resulting formulas express the cross
ratios $u_{as}$ through the primed system parameters and a solution
of the Eqs.\ \eqref{TBABA}. We can solve these equations for the
primed system parameters in terms of the parameters at the starting
point of the continuation. Of course, the relation depends on the
choice of a solution to the Bethe Ansatz Eqs.\  \eqref{TBABA}.
\par
Let us finally discuss the form of the free energy \eqref{eq:afree}.
Upon analytic continuation of the parameters, the Y-functions may
again give rise to pole terms that cross the real axis, because the
integrand has the same structure as in the Y-system. This happens
precisely when the conditions \eqref{poleeq} are satisfied. After
taking the multi-Regge limit, only these pole terms survive and
we obtain an expression of the form
\begin{equation}
  A'_\mathrm{free}  =  - \sum_{a,s} \sum_{\nu=1}^{N_{a,s}}
i(\sqrt{2})^{\delta_{a,2}} \mu^{(a,s)}_\nu |m_s|'
\sinh(\theta^{(a,s)}_\nu).
\label{eq:AfreeMRL}
\end{equation}
Recall that the quantities $m'_s$ and $\phi'_s$ are considered as
functions of the cross ratios and the Bethe roots $\theta^{(a,s)}_\nu$,
as described in the previous paragraph. Once the explicit expressions
are plugged in, we can extract the leading terms in the multi Regge
limit $|m_s|'\rightarrow\infty$. This completes our task to explain
the role of the Bethe Ansatz \eqref{TBABA} in the multi-Regge regime
of strongly coupled $\mathcal{N}=4$ SYM theory. \par
We have shown that upon analytic continuation of the cross
ratios, and hence of the system parameters $|m_s|,C_s$ and $\phi_s$, the
Y-system can pick up additional terms from residue contributions. These
may be thought off as excitations above the ground state. As we approach
the multi-Regge regime, i.e.\ send the mass parameters $m_s$ to infinity,
the quantum fluctuations are suppressed, as explained before. In this limit, the free energy of
the system is determined by the bare energies of the excitations that
were produced during the continuation, as expressed in Eq. \eqref{eq:AfreeMRL}. In case no excitations appeared, the free energy
would continue to vanish in the multi-Regge limit. We shall, however, see excitations for most of the regions we analyze below.

In order to conclude this section and warm up for the next, let us illustrate
all the above for the example of $n=6$ external gluons, see \cite{Bartels:2010ej,
Bartels:2013dja}. Since for six gluons the parameter $s$ is fixed to $1$, we
will suppress it in the following.
In \cite{Bartels:2010ej, Bartels:2013dja}, we start from a set of parameters
$(|m|,C,\phi)$ with $\phi$ small and continue along a particular curve such
that the cross ratio $u_1$ performs a full rotation, while the other cross
ratios remain fixed, see path $P_{6,--}$ in subsection \ref{sec:mrl_regions}. At the endpoint
of the continuation we reach another point $(|m|',C',\phi')$ in the
complexified parameter space corresponding to the same values of the cross
ratios, i.e.\ $u_a (|m|,C,\phi) = u_a(|m|',C',\phi')$. A numerical analysis
of the Y-system equations shows that, along this path, two solutions of the
equation $\tilde{\Yf}_3(\theta_\ast) =-1$  (or of $\tilde{\Yf}_1(\theta_\ast)$,
depending on the value of $C$) cross the real axis while two solutions of $\tilde{\Yf}_2(\theta_\ast)=-1$ approach the real axis.
The solution of $\tilde{\Yf}_3(\theta_\ast)=-1$ crossing into the positive
(negative) half-plane will be called $\theta_+$ ($\theta_-$) in the
following. For the $6$-gluon case, there is a special symmetry of the
Y-system,
\begin{equation}
  \tilde{\Yf}_{a}(\theta)=\tilde{\Yf}_{a}(-\theta).
  \label{eq:ysym_6pt}
\end{equation}
In \cite{Bartels:2013dja} we show that due to this symmetry, the two solutions of $\tilde{\Yf}_2$ pinch the integration contour, but never cross and therefore do not contribute to the free energy term or the Y-system. We can then spell out the equations at the
endpoint of the continuation,
\begin{align}
  \log \tilde{\Yf}_2(\theta) &=-\sqrt{2}|m|'\cosh\theta +\log S_2(\theta-\theta_-) -\log S_2(\theta-\theta_+),\label{eq:6pt_ysys_endp2}\\
  \log \tilde{\Yf}_{2\pm 1}(\theta)&=-|m|'\cosh\theta \pm C'+\log S_1(\theta-\theta_-) -\log S_1(\theta-\theta_+)\ .
  \label{eq:6pt_ysys_endp31}
\end{align}
Since there are two Bethe roots, the Bethe Ansatz consists of two equations. It turns
out that these equations are identical due to the above-mentioned symmetry so that we end up
with
\begin{equation}
  -e^{p'_3(\theta_+)}  = \frac{S_1(\theta_+-\theta_-)}{S_1(0)},
      \label{BAhex2}
\end{equation}
where $S_1(\theta)$ was spelled out in Eq.\  (\ref{eq:basic_smat}) and where $p'_3(\theta_+)=|m|'\cosh\theta_+$.
Note that $\mu^{(3)}_- = - \mu^{(3)}_+ = 1$.
When we perform the multi-Regge limit $|m|'\rightarrow\infty$ in Eq.\  (\ref{BAhex2}), the
left hand side of our Bethe Ansatz equation diverges and hence
$\theta_+-\theta_-$ has to approach a pole of $S_1$. Hence, because of the pole structure of $S_1(\theta)$, our solution consists of one Bethe string
with
\begin{equation}
  \theta_+-\theta_-=i\frac{\pi}{2}.
\end{equation}
Furthermore, taking into account the symmetry Eq.\  (\ref{eq:ysym_6pt}) which relates the endpoints $\theta_\pm$ as
\begin{equation}
  \theta_-=-\theta_+
\end{equation}
we can conclude that the endpoints of the crossed solutions are at
\begin{equation}
  \theta_\pm=\pm i\frac{\pi}{4}.
  \label{eq:6pt_endpoints}
\end{equation}
Using Eqs.\  (\ref{eq:6pt_ysys_endp2}),(\ref{eq:6pt_ysys_endp31}) together with the recursion relations, we obtain the cross ratios
\begin{align}
  u'_1&=1 -\gamma\varepsilon'\left( w^\prime+\frac{1}{w'}-2 \cosh C^\prime\right),\\
u'_2&= \gamma w^\prime\varepsilon^\prime,\\
u'_3&= \gamma \frac{\varepsilon^\prime}{w'},
\end{align}
with $\gamma = - (3 + 2 \sqrt 2)$ and the parameters $w^\prime,\varepsilon^\prime$ are related to the primed parameters through Eqs.\ (\ref{eq:params_eps_w}).
The above expressions are valid up to corrections of $\mathcal{O}(\varepsilon^{\prime 2})$.
Imposing $u_a=u'_a$ we obtain relations between the primed and the unprimed parameters and find
\begin{equation}
  \varepsilon^\prime=\frac{1}{\gamma}\varepsilon,\quad w^\prime=w,\quad \cosh C^\prime=-\cosh C, \label{wecprime}
\end{equation}
again up to subleading corrections.
Now we can compute the free energy in terms of the primed parameters
through Eq.\  (\ref{eq:AfreeMRL}) and then express the primed parameters
through the original ones to find
\begin{equation}
A^{\prime(6)}_{\textrm{free}} = \sqrt{2}\log\varepsilon - \sqrt{2}\log \gamma +
O\left(\varepsilon\log\varepsilon\right).
\end{equation}
This ends our example, for more details on the calculation see \cite{Bartels:2013dja}.
We can perform a similar analysis for any solution of the Bethe Ansatz Eq.\ \eqref{TBABA} and will do so for the 7-point amplitude in the next section.

\section{Results for the $7$-point amplitude}
\label{sec:calcamp}
We now turn to the explicit calculation of the remainder function in various Regge regions of the $7$-point amplitude. The kinematics and the various non-trivial multi-Regge regions for this amplitude were described in section \ref{sec:ysys_mrl}. In principle it is not
difficult to follow our general algorithm for the calculation of the remainder function
at strong coupling. We know that each region corresponds to a particular pattern of
Bethe roots. Once these are determined as a solution of Eqs.\ \eqref{TBABA}, calculating
the remainder function is a bit cumbersome, but straightforward. The main issue is to
associate a solution of Eq.\ \eqref{TBABA} with each of the four non-trivial multi-Regge
regions that exist for $n=7$. So far, the only way we can make this association is
through a numerical analysis of the Y-system. Once we have understood which solutions
of the Eqs.\ \eqref{poleeq} cross the real axis, the relevant solution and the
amplitude can be computed analytically.

For the numerical investigations it is advantageous to reformulate the original
Y-system such that the driving terms contain the cross ratios rather than the
system parameters $m_s,C_s$ and $\phi_s$. This will be explained in the first
subsection. Then we discuss each of the four non-trivial multi-Regge regions,
starting with the region $(--+)$. In each case we present our numerical results
before we apply the insights they provide to calculate the remainder functions.
We also compare the final expressions for the remainder functions with the
expectations from weak coupling.

\subsection{An alternative Y-system}
\label{sec:alt_ysys}
As described in the general algorithm in section \ref{sec:ba}, a crucial part
of the calculation is to follow the solutions of the equations
\begin{equation}
  \tilde{\Yf}_{a,s}(\theta)=-1
\end{equation}
through the $\theta$-plane as we analytically continue the cross ratios. At the
moment we can only do this through numerical studies of the Y-system. But with
the Y-system (\ref{eq:ysys1})-(\ref{eq:ysys3}) this is a rather difficult task
as we would first need to determine how the auxiliary parameters $|m_s|$, $C_s$
and $\phi_s$ behave along the path of continuation. In order to circumvent the
issue, we shall pass to a different version of the Y-system, similar to the one
derived in appendix F of \cite{Alday:2010ku}.\par
Note that our choice of cross ratios can be obtained via the recursion relations if we know the
following special values of Y-functions
\begin{equation}
  \tilde{\Yf}_{1,1}(0),\,\tilde{\Yf}_{2,1}\left(-i\frac{\pi}{4}\right),\,
  \tilde{\Yf}_{3,1}(0),\,\tilde{\Yf}_{1,2}(0),\,\tilde{\Yf}_{2,2}
  \left(i\frac{\pi}{4}\right),\,\tilde{\Yf}_{3,2}(0),
  \label{eq:newvar}
\end{equation}
Here, all arguments lie in the fundamental strip $|\mathrm{Im}\,\theta|\leq\frac{\pi}{4}$. In deriving concrete
expressions one must use our choice of phases Eq.\  (\ref{eq:ysys_params_mrl})
to relate cross ratios defined in terms of Y-functions to the $\tilde{\Yf}$-functions.
Since we prescribe the behavior of the cross ratios, we can use the recursion relations
to infer the behavior of the values listed in Eq.\ (\ref{eq:newvar}) during the analytic continuation. Consequently, these values of our Y-functions are better suited as parameters in the Y-system equations than the auxiliary parameters $|m_s|$, $C_s$ and $\phi_s$.
In order to work out the relevant equations, we solve the Y-system equations at the specific values of the spectral parameter $\theta$ that appear in Eq.\ (\ref{eq:newvar}) for the auxiliary parameters to obtain
\begin{equation}
\begin{aligned}
C_s=&\frac{1}{2}\log\left(\frac{\tilde{\Yf}_{3,s}(0)}{\tilde{\Yf}_{1,s}(0)}\right)-\frac{1}{2}K_3\star\gamma_s\big|_{\theta=0},\\
\left|m_s\right|=&-\frac{1}{2}\log\left(\tilde{\Yf}_{1,s}(0)\tilde{\Yf}_{3,s}(0)\right)-\frac{1}{2}K_2\star\beta_s\big|_{\theta=0}-K_1\star\alpha_s\big|_{\theta=0}.
\end{aligned}
\label{eq:auxparams}
\end{equation}
Note that none of the $\tilde{\Yf}_{2,s}$ variables appears in Eqs.\ (\ref{eq:auxparams}).
The reason for this is that we are working with fixed phases for simplicity and therefore do not replace the parameters $\phi_s$ in the original equations\footnote{Note that for some of the paths describes in section \ref{sec:ysys_mrl} we have a relative rotation of small cross ratios which entails that we cannot keep the phases $\phi_s$ fixed during the continuation.
However, we can estimate the size of the error by adding a small deviation from the fixed value, $\phi_s=(1-s)\frac{\pi}{4}+\epsilon_\phi$. Then, a relative rotation of $\pm 2i\varphi$ would entail
  \begin{equation}
    \frac{u_{2s}}{u_{3s}}=w_s^{2}=e^{2|m_s| \sin(\phi_s-(1-s)\frac{\pi}{4})}=e^{2|m_s|\epsilon_\phi}\stackrel{!}{=}e^{\pm 2i\varphi},
  \end{equation}
  which gives
  \begin{equation}
    \epsilon_\phi=\frac{\pm i\varphi}{|m_s|}\sim\mathcal{O}\left( \frac{1}{\log\varepsilon_s} \right)
    \label{eq:error_fixedphi}
  \end{equation}
  and is therefore negligible throughout the continuation.}.
We can now plug Eqs.\ (\ref{eq:auxparams}) in the original Y-system equations and obtain the new Y-system
\begin{align}
  \nonumber \log\tilde{\Yf}_{1,s}(\theta)=&\frac{1}{2}\log\left(\tilde{\Yf}_{1,s}(0)\tilde{\Yf}_{3,s}(0)\right)
  \cosh\theta-\frac{1}{2}\log\left(\frac{\tilde{\Yf}_{3,s}(0)}{\tilde{\Yf}_{1,s}(0)}\right)\\[2mm]
  &\quad +\sum\limits_{a',s'}\int d\theta' \mathcal{K}_{s,s'}^{1,a'}(\theta,\theta')\log\Yint{a'}{s'},\label{eq:newysys1} \\[2mm]
  \log\tilde{\Yf}_{2,s}(\theta)=&\frac{1}{\sqrt{2}}
  \log\left(\tilde{\Yf}_{1,s}(0)\tilde{\Yf}_{3,s}(0)\right) \cosh\theta +\sum\limits_{a',s'}\int d\theta' \mathcal{K}_{s,s'}^{2,a'}(\theta,\theta')\log\Yint{a'}{s'},\label{eq:newysys2}\\[2mm]
 \nonumber  \log\tilde{\Yf}_{3,s}(\theta)=&\frac{1}{2}\log\left(\tilde{\Yf}_{1,s}(0)
 \tilde{\Yf}_{3,s}(0)\right)\cosh\theta+
 \frac{1}{2}\log\left(\frac{\tilde{\Yf}_{3,s}(0)}{\tilde{\Yf}_{1,s}(0)}\right)\\[2mm]
 &\quad + \sum\limits_{a',s'}\int d\theta' \mathcal{K}_{s,s'}^{3,a'}(\theta,\theta')\log\Yint{a'}{s'}.\label{eq:newysys3}
\end{align}
The new kernels are more complicated compared to the kernels of the original Y-system and are spelled out in appendix \ref{sec:app_kernels}.
\subsection{Remainder function in the Regge region $(--+)$}
\label{sec:region_mmp}
\begin{figure}[t]
  \centering
  \includegraphics[scale=.75]{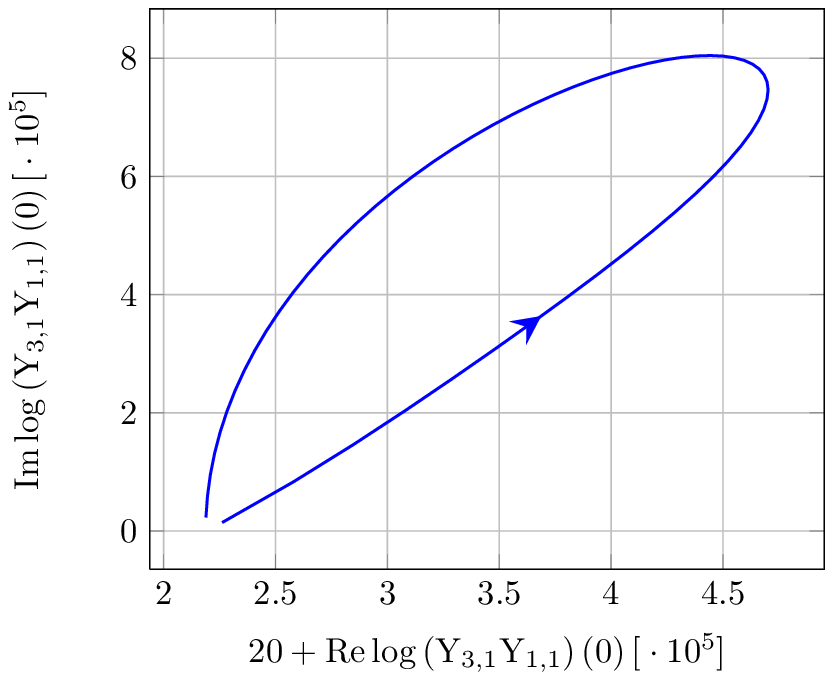}
  \includegraphics[scale=.75]{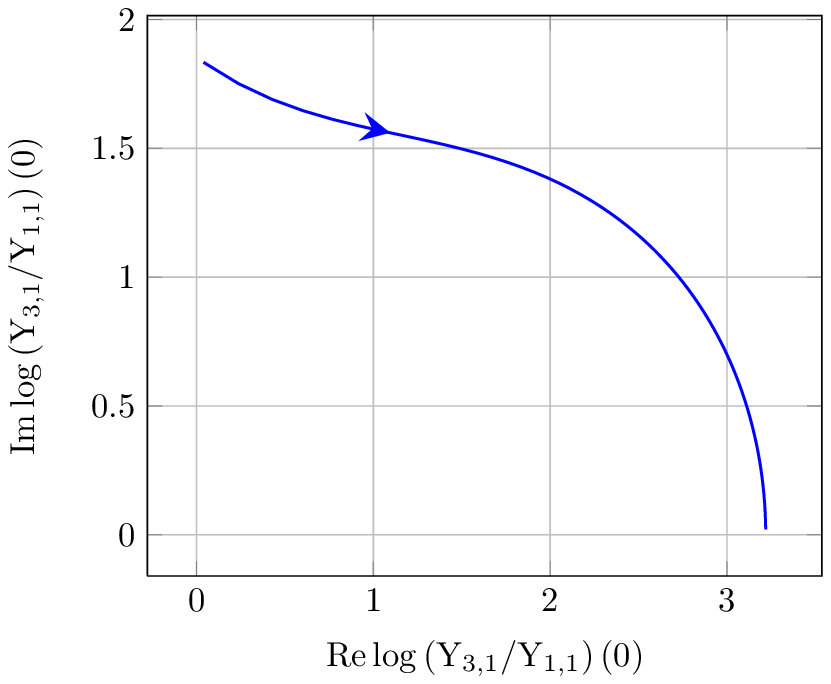}
  \includegraphics[scale=.75]{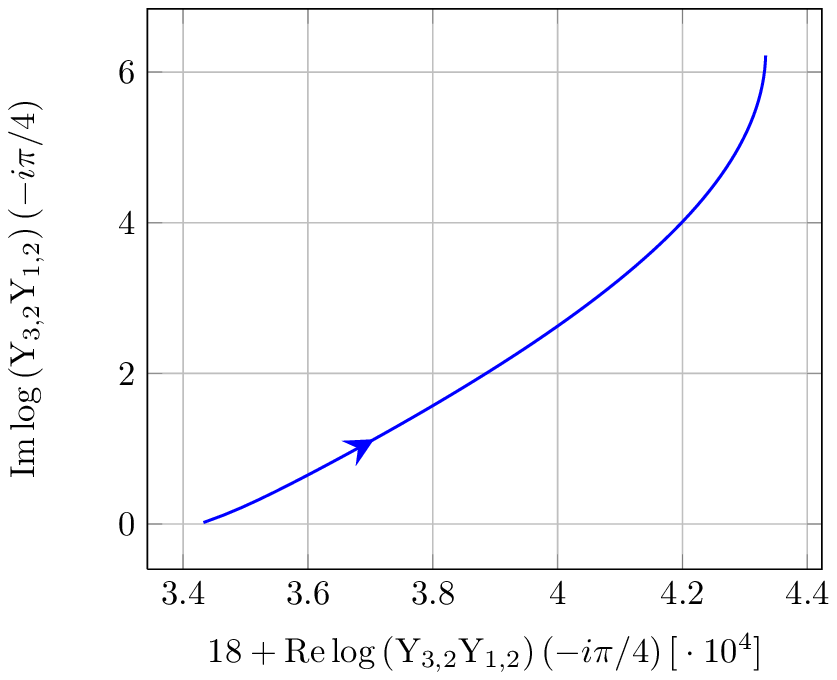}
  \includegraphics[scale=.75]{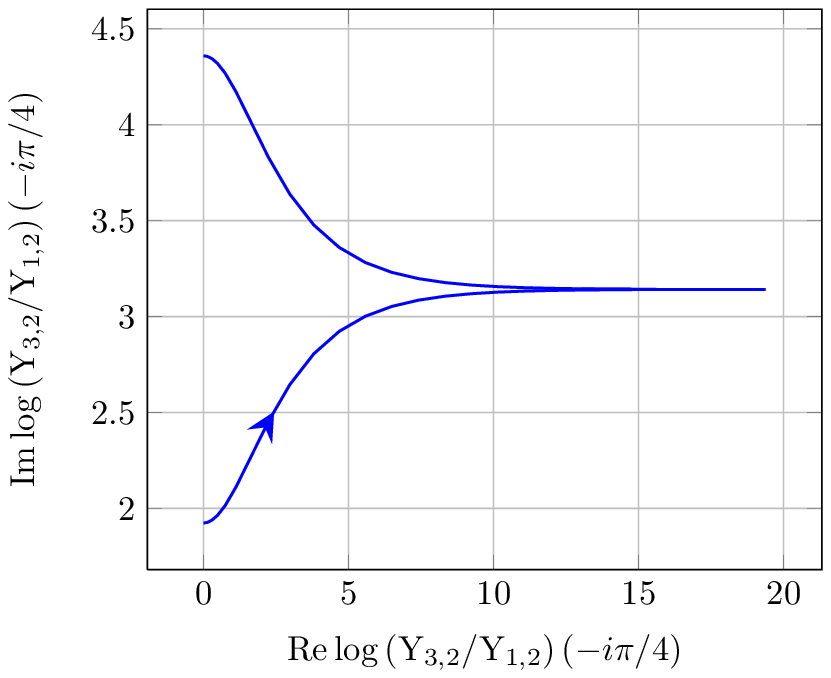}
  \caption{Paths of the driving terms during the analytic continuation for the path Eq.\  (\ref{eq:7pt_mmp}). The starting values for the parameters chosen here are $|m_1|=10$, $|m_2|=9$, $C_1=\mathrm{arccosh}\left(\frac{3}{5}\right)$, $C_2=\mathrm{arccosh}\left(\frac{4}{7}\right)$. Note that some axes have been shifted and rescaled. The direction of growing $\varphi$ is indicated by the arrows.}
  \label{fig:P4DrvTerms}
\end{figure}

We now evaluate the remainder function in the first relevant Regge region, namely the region $(--+)$, which we probe via the path $P_{7,--+}$ that was defined in Eq.\ \eqref{eq:7pt_mmp}. Prescribing this behavior of the cross ratios, we solve the recursion relations for the driving terms of the modified Y-system Eqs.\  (\ref{eq:newysys1})-(\ref{eq:newysys3}) numerically.
We find the results shown in figure \ref{fig:P4DrvTerms}.
\subsubsection{Numerical analysis of the continuation}
After finding the paths of continuation for the driving terms, we now turn to the determination of $\mathrm{A}_{\mathrm{free}}$.
To do so, we need to determine those $\tilde{\Yf}_{a,s}$-functions for which a solution to the equation $\tilde{\Yf}_{a,s}=-1$ crosses the real axis, as explained in detail in section \ref{sec:ba}.
Since we know the behavior of the system parameters in the new Y-system along the path of continuation, we can simply solve the Y-system in the complex $\theta$-plane for each value of $\varphi$ and then determine the location of the solutions $\tilde{\Yf}_{a,s}=-1$.
We solve the Y-system by an iterative procedure, similar to the algorithm presented in \cite{Alday:2010vh}.\par
\begin{figure}[t]
  \centering
  \includegraphics[scale=.75]{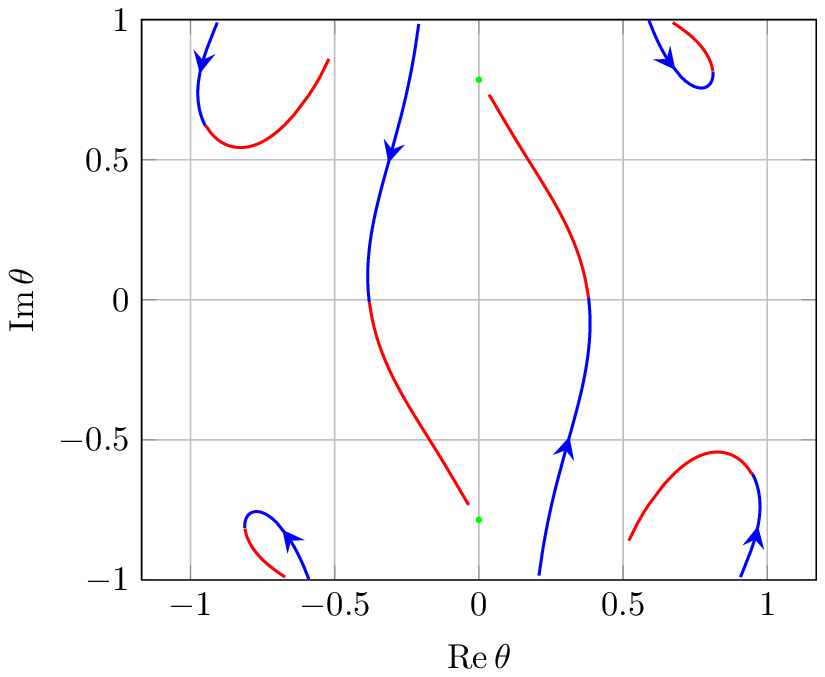}
  \includegraphics[scale=.75]{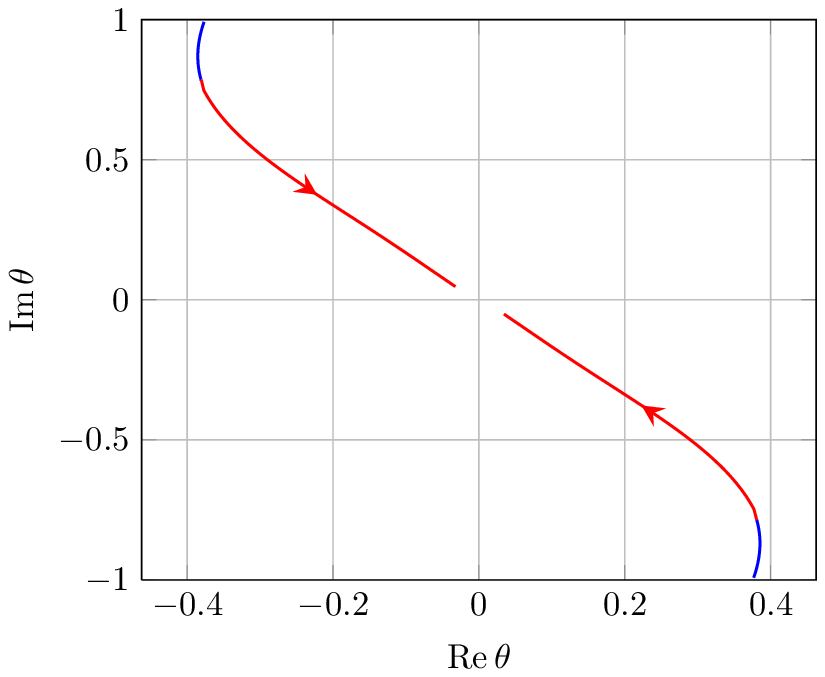}
  \caption{Left: Crossing solutions of $\tilde{\Yf}_{3,2}(\theta)=-1$ during the continuation Eq.\  (\ref{eq:7pt_mmp}). We find that two solutions cross the real axis and approach the endpoints $\pm i\frac{\pi}{4}$. Right: Towards the end of the continuation, a pair of solutions of $\tilde{\Yf}_{2,2}(\theta)=-1$ approaches the real axis, but does not contribute to the remainder function as is argued in the main text. The direction of growing $\varphi$ is indicated by the arrows.}
  \label{fig:7pt_p4_crossing}
\end{figure}

Following this procedure, we find the behavior for the $\tilde{\Yf}$-functions displayed in figure \ref{fig:7pt_p4_crossing}.
We see that a pair of solutions of $\tilde{\Yf}_{3,2}$ crosses the real axis and approaches the values $\pm i \frac{\pi}{4}$ at the end of the continuation.
Furthermore, two solutions of $\tilde{\Yf}_{2,2}$ approach the origin at the end of the continuation.
As we show in the next section, the Y-system equations for the triplet $\tilde{\Yf}_{a,2}$ at the end of the continuation are of the same form as the equations for the $6$-point case in section \ref{sec:ba}.
Therefore, we can argue as in the $6$-point case that the pair of solutions of $\tilde{\Yf}_{2,2}$ never crosses the integration contour and we just have to consider the pair of crossing solutions of $\tilde{\Yf}_{3,2}(\theta)$.
Furthermore, since we find the same equations as in the $6$-point case for the triplet $\tilde{\Yf}_{a,2}$, we can use the endpoint conditions Eq.(\ref{TBABA}) to analytically determine the endpoint of the Bethe roots to be $\pm i\frac{\pi}{4}$, confirming our numerical analysis.

\subsubsection{Calculation of the remainder function}

Using the input from the numerical analysis, we can now determine the amplitude.
Let us stress again that the input from the numerics is discrete - it only provides information on which Y-functions have crossed.
The endpoints can be determined analytically, therefore our result derived below is not bound to any numerical accuracy.
Note also that since we are done with the numerical analysis, we can switch back to the standard Y-system Eqs.\ (\ref{eq:ysys1})-(\ref{eq:ysys3}), because the crossing solutions of the $\tilde{\Yf}$-functions are independent of the choice of system parameters, of course.
As a first step, let us determine the cross ratios at the end of the continuation.
In the following, all quantities that have been analytically continued are marked with a prime, as before.
At the end of the continuation, the Y-system takes the following form:
\begin{align}
  \nonumber\log\tilde{\Yf}'_{1,s}&=-|m_s|'\cosh\theta-C'_s+\sum\limits_{a',s'}\int d\theta' K_{s,s'}^{1,a'}\left(\theta-\theta'+i\phi'_s-i\phi'_{s'}\right)\log\left(1+\tilde{\Yf}'_{a',s'}(\theta')\right)\\
  &+\log \frac{S_{s,2}^{1,3}(\theta+i\frac{\pi}{4}+i\phi'_s-i\phi'_2)}{S_{s,2}^{1,3}(\theta-i\frac{\pi}{4}+i\phi'_s-i\phi'_2)},\label{eq:ysys1ac}\\
 \nonumber\log\tilde{\Yf}'_{2,s}&=-\sqrt{2}|m_s|'\cosh\theta+\sum\limits_{a',s'}\int d\theta' K_{s,s'}^{2,a'}\left(\theta-\theta'+i\phi'_s-i\phi'_{s'}\right)\log\left(1+\tilde{\Yf}'_{a',s'}(\theta')\right)\\
  &+\log \frac{S_{s,2}^{2,3}(\theta+i\frac{\pi}{4}+i\phi'_s-i\phi'_2)}{S_{s,2}^{2,3}(\theta-i\frac{\pi}{4}+i\phi'_s-i\phi'_2)},\label{eq:ysys2ac}\\
  \nonumber\log\tilde{\Yf}'_{3,s}&=-|m_s|'\cosh\theta+C'_s+\sum\limits_{a',s'}\int d\theta' K_{s,s'}^{3,a'}\left(\theta-\theta'+i\phi'_s-i\phi'_{s'}\right)\log\left(1+\tilde{\Yf}'_{a',s'}(\theta')\right)\\
  &+\log \frac{S_{s,2}^{3,3}(\theta+i\frac{\pi}{4}+i\phi'_s-i\phi'_2)}{S_{s,2}^{3,3}(\theta-i\frac{\pi}{4}+i\phi'_s-i\phi'_2)}.\label{eq:ysys3ac}
\end{align}
As we now go to the multi-Regge limit, we can neglect the integrals and are left with the equations
\begin{align}
  \Yf'_{1,s}&=\left(e^{-|m_s|'\cosh(\theta-i\phi'_s)-C'_s}\right)\frac{S_{s,2}^{1,3}(\theta+i\frac{\pi}{4}-i\phi'_2)}{S_{s,2}^{1,3}(\theta-i\frac{\pi}{4}-i\phi'_2)},\\
  \Yf'_{2,s}&=\left(e^{-\sqrt{2}|m_s|'\cosh(\theta-i\phi'_s)}\right)\frac{S_{s,2}^{2,3}(\theta+i\frac{\pi}{4}-i\phi'_2)}{S_{s,2}^{2,3}(\theta-i\frac{\pi}{4}-i\phi'_2)},\\
  \Yf'_{3,s}&=\left(e^{-|m_s|'\cosh(\theta-i\phi'_s)+C'_s}\right)\frac{S_{s,2}^{3,3}(\theta+i\frac{\pi}{4}-i\phi'_2)}{S_{s,2}^{3,3}(\theta-i\frac{\pi}{4}-i\phi'_2)}.
\end{align}
Using the recursion relations and introducing the parameters
\begin{equation}
  \varepsilon':=e^{-|m_s|'\cos\left( (s-1)\frac{\pi}{4}+\phi'_s\right)},\, w':=e^{|m_s|'\sin\left( (s-1)\frac{\pi}{4}+\phi'_s\right)},
\end{equation}
we find that the cross ratios at the end of the path are given by\footnote{Note that in evaluating the cross ratios we set $\phi'_s=\phi_s$ in the S-matrix factors. The error we are making with this choice is of the same order as in Eq.\  (\ref{eq:error_fixedphi}) and therefore negligible.}
\begin{equation}
\begin{alignedat}{3}
  u'_{11}&=1-\gamma\varepsilon'_2\left(w'_2+\frac{1}{w'_2}-2\cosh C'_2\right),\quad && u'_{21}=\gamma w'_2\varepsilon'_2,\quad && u'_{31}=\gamma\frac{\varepsilon'_2}{w'_2},\\
  u'_{12}&=1+\varepsilon'_1\left(\frac{1}{\gamma}w'_1+\frac{1}{w'_1}+\frac{2}{\sqrt{-\gamma}}\sinh C'_1\right),\quad && u'_{22}=-\frac{1}{\gamma}w'_1\varepsilon'_1,\quad && u'_{32}=-\frac{\varepsilon'_1}{w'_1},
\end{alignedat}
  \label{eq:crsac}
\end{equation}
where $\gamma=-3-2\sqrt{2}$, and all above expressions are valid up to corrections of $\mathcal{O}(\varepsilon'^2)$.
For the path under consideration we then impose $u'_{22}=-u_{22}$, $u'_{32}=-u_{32}$, as well as $u'_{as}=u_{as}$ for all other cross ratios.
This determines $\varepsilon'_s$, $w'_s$ and $C'_s$ in terms of the old parameters, giving rise to the relations
\begin{equation}
\begin{alignedat}{3}
\varepsilon'_1&=\sqrt{\gamma}\varepsilon_1,\quad && w'_1=\sqrt{\gamma}w_1, \quad && \cosh C'_1=\sqrt{1-\left(w_1+w_1^{-1}+\cosh C_1\right)^2},\\
  \varepsilon'_2&=\frac{1}{\gamma}\varepsilon_2,\quad && w'_2=w_2,\quad && \cosh C'_2=-\cosh C_2,
\end{alignedat}
  \label{eq:epsac}
\end{equation}
again up to corrections $\mathcal{O}(\varepsilon^2)$.
We can now examine the terms of the remainder function Eq.\ (\ref{eq:def_remainder_fct})
after the continuation piece by piece and see how they contribute. We start with $\mathrm{A}_{\mathrm{per}}$, which after the continuation reads
\begin{equation}
\begin{aligned}
\mathrm{A}'_{\mathrm{per}}=&\frac{1}{2}\left(\log^2\varepsilon'_1+\log^2 w'_1+\log^2\varepsilon'_2+\log^2 w'_2\right.\\
&\quad\left.+\log\varepsilon'_1\log\varepsilon'_2+\log w'_1\log w'_2+\log\varepsilon'_2\log w'_1-\log\varepsilon'_1\log w'_2\right).
\end{aligned}
  \label{eq:aperac}
\end{equation}
Using Eq.\  (\ref{eq:epsac}), this gives
\begin{equation}
  \mathrm{A}'_{\mathrm{per}}=\mathrm{A}_{\mathrm{per}}+\frac{1}{4}\log^2\gamma-\frac{1}{2}\log\gamma\log\varepsilon_2.
  \label{eq:aperdiff}
\end{equation}
We next turn to $\mathrm{A}_{\mathrm{free}}$.
We determined the relevant crossings already in the last section.
Furthermore, we know from section \ref{sec:ba} that in the multi Regge-limit only the residue terms from the crossing $\tilde{\Yf}$-functions contribute to the free energy.
In the last section we saw that one pair of solutions of $\tilde{\Yf}_{3,2}$ crosses the real axis and approaches the endpoints $\pm i \frac{\pi}{4}$.
After neglecting the integrals, we are therefore left with
\begin{align}
\nonumber  \mathrm{A}'_{\mathrm{free}}=&-\frac{|m_2|}{2\pi}\left(2\pi i\sinh\left(-i\frac{\pi}{4}\right)-2\pi i\sinh\left(i\frac{\pi}{4}\right)\right)\\
=&-\sqrt{2}\sqrt{\log^2\varepsilon'_2+\log^2 w'_2}\label{eq:afreeac}\\
\nonumber \approx& +\sqrt{2}\log\varepsilon'_2=+\sqrt{2}\left(\log\varepsilon_2-\log\gamma\right).
\end{align}
This leaves us with $\Delta'$, whose form before continuation was spelled out in Eq.\  (\ref{eq:deltabc}).
During the continuation, some cut contributions have to be picked up and we end up with
\begin{align}
\nonumber
  \Delta'=&\Delta-i\frac{\pi}{4}\left(2\log u_{11} +\log u_{12}+\log \tilde{u}-2\log(1-u_{11})\right)\\
  &+\frac{1}{4}\left(\mathrm{Li}_2(1-u_{22})-\mathrm{Li}_2(1+u_{22})+\mathrm{Li}_2(1-u_{32})-\mathrm{Li}_2(1+u_{32})\right)+\mathrm{const.}\label{eq:deltaac}\\
\nonumber  =&-i\frac{\pi}{2}\log\varepsilon_2-i\frac{\pi}{2}\log\left(w_2+w_2^{-1}+2\cosh C_2\right)+\mathrm{const.}+\mathcal{O}(\varepsilon).
\end{align}
Adding all contributions and using\footnote{One way of fixing the ambiguity of the imaginary part of $\log\gamma$ is to calculate the original Y-system parameter $|m_s|$ numerically via Eq.\  (\ref{eq:auxparams}) during the continuation. Comparing the numerical value at the endpoint with the analytic value for $|m_s|'$ obtained from Eq.\  (\ref{eq:epsac}) fixes $\log\gamma$.} $\log\gamma=\log|\gamma|-i\pi$ we find
\begin{equation}
  A'_{\mathrm{per}}+A'_{\mathrm{free}}+\Delta'+i\delta'_{7,--+}=-e_2\log\varepsilon_2-i\pi e_2+\mathrm{const.}+\mathcal{O}(\varepsilon),
  \label{eq:remac}
\end{equation}
where
\begin{equation}
  e_2=-\sqrt{2}+\log\left(1+\sqrt{2}\right)
  \label{eq:e2}
\end{equation}
and where
\begin{equation}
  \delta'_{7,--+}=\frac{\pi}{2}\log\left(w_2+\frac{1}{w_2}+2\cosh C_2 \right)
  \label{eq:delta_phase_p4}
\end{equation}
is a phase that cancels with an equivalent term in the BDS-part in the full amplitude.
To rewrite this result in terms of the cross ratios we use the relations
\begin{equation}
  \varepsilon_2=\sqrt{\tilde{u}_{21}\tilde{u}_{31}}(1-u_{11}),\quad w_2=\sqrt{\frac{u_{21}}{u_{31}}}
  \label{eq:relparamcrs}
\end{equation}
and obtain
\begin{equation}
  A'_{\mathrm{per}}+A'_{\mathrm{free}}+\Delta'+i\delta'_{7,--+}=\frac{\sqrt{\lambda}}{2\pi}\left(e_2\log\left(-(1-u_{11})\tilde{u}_{21}\tilde{u}_{31}\right)+\mathrm{const.}\right).
  \label{eq:raccrs}
\end{equation}
Exponentiating, we obtain our final result for the remainder function
\begin{equation}
  \left.e^{R_{7,--+}+i\delta_{7,--+}}\right|_{\mathrm{MRL}}\sim \left(-(1-u_{11})\sqrt{\tilde{u}_{21}\tilde{u}_{31}}\right)^{\frac{\sqrt{\lambda}}{2\pi}e_2},
  \label{eq:rem_fct_p4}
\end{equation}
where
\begin{equation}
  \delta_{7,--+}=\frac{\sqrt{\lambda}}{2}\log\left(\tilde{u}_{21}\tilde{u}_{31}\right)=\frac{\pi}{2}\gamma_K\log\left(\tilde{u}_{21}\tilde{u}_{31}\right)
  \label{eq:fullphase_mmp}
\end{equation}
with the strong coupling limit of the cusp anomalous dimension $\gamma_K=\frac{\sqrt{\lambda}}{\pi}$.
Remarkably, this result can be expressed in the same form as the $6$-point result, $R_{7,--+}(u_{as})=R_{6,--}(u_{11},u_{21},u_{31})$, and nicely matches the weak-coupling predictions as stated in section \ref{sec:wc}.\par
Note that in this section we have focused on the calculation for the region $P_{7,--+}$.
The calculation for the region $P_{7,+--}$ is very similar and is therefore presented in appendix \ref{sec:region_pmm}.
Here we only present the final result, which reads
\begin{equation}
  e^{R_{7,+--}+i\delta_{7,+--}}\sim \left(-(1-u_{12})\sqrt{\tilde{u}_{22}\tilde{u}_{32}}\right)^{\frac{\sqrt{\lambda}}{2\pi}e_2},
  \label{eq:rem_fct_p5}
\end{equation}
with
\begin{equation}
  \delta_{7,+--}=\frac{\pi}{2}\gamma_K\log\left(\tilde{u}_{22}\tilde{u}_{32}\right)
  \label{eq:fullphase_mmp}
\end{equation}
showing manifestly the target-projectile symmetry Eq.\  (\ref{eq:target_projectile}) we have mentioned before.

\subsection{Remainder function in the Regge region $(---)$}
\begin{figure}[t]
  \centering
  \includegraphics[scale=.75]{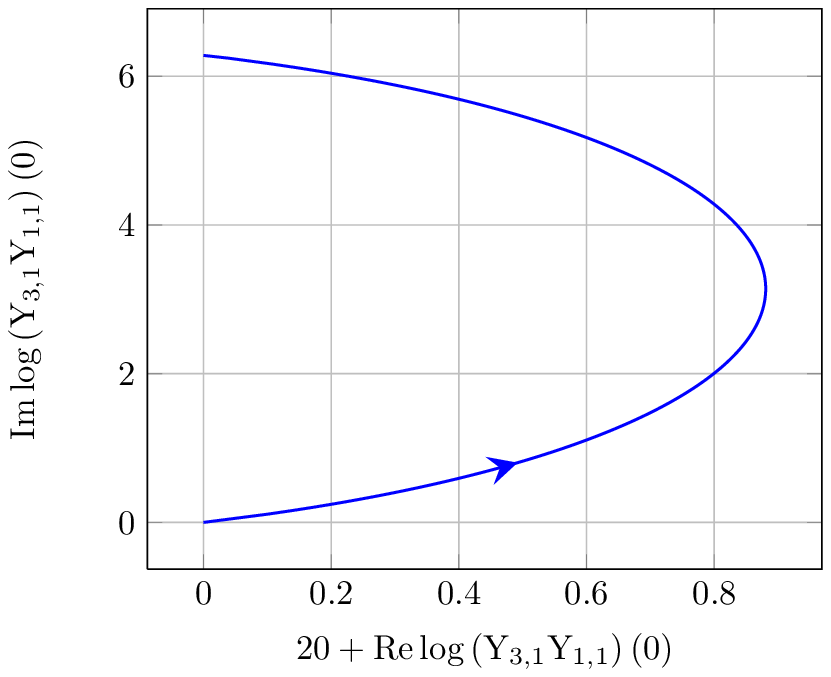}
  \includegraphics[scale=.75]{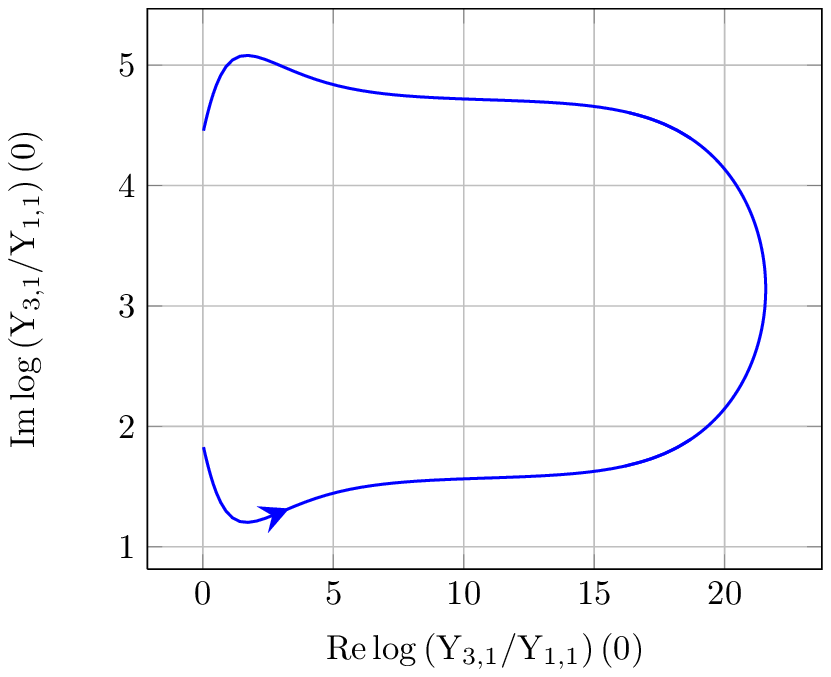}
  \caption{Solution of the driving terms during the continuation Eq.\  (\ref{eq:7pt_mmm}). The direction of growing $\varphi$ is indicated by the arrows. For simplicity, the sets of parameters are identified as $|m|=|m_1|=|m_2|=10$, $C=C_1=-C_2=\mathrm{arccosh}\left(\frac{3}{5}\right)$ at the starting point.}
  \label{fig:P3DrvTerms}
\end{figure}

After having discussed the paths with only two flipped legs, we now turn to the path where all produced particles are chosen to be incoming. As explained in section \ref{sec:kinematics}, the naive continuation along (semi-)circles fails for this path as
it does not satisfy the conformal Gram relation. An appropriate deformation that is
consistent with the Gram relation was spelled out in Eq.\ \eqref{eq:7pt_mmm} and this
is the one we will be using throughout our analysis.
As before, we use the recursion relations to determine the behavior of the driving terms of the modified Y-system Eqs.\  (\ref{eq:newysys1})-(\ref{eq:newysys3}).
A plot of the curves followed by the driving terms during the continuation is shown in figure \ref{fig:P3DrvTerms}.
\subsubsection{Numerical analysis of the continuation}
\begin{figure}
	\centering
	\includegraphics[scale=.8]{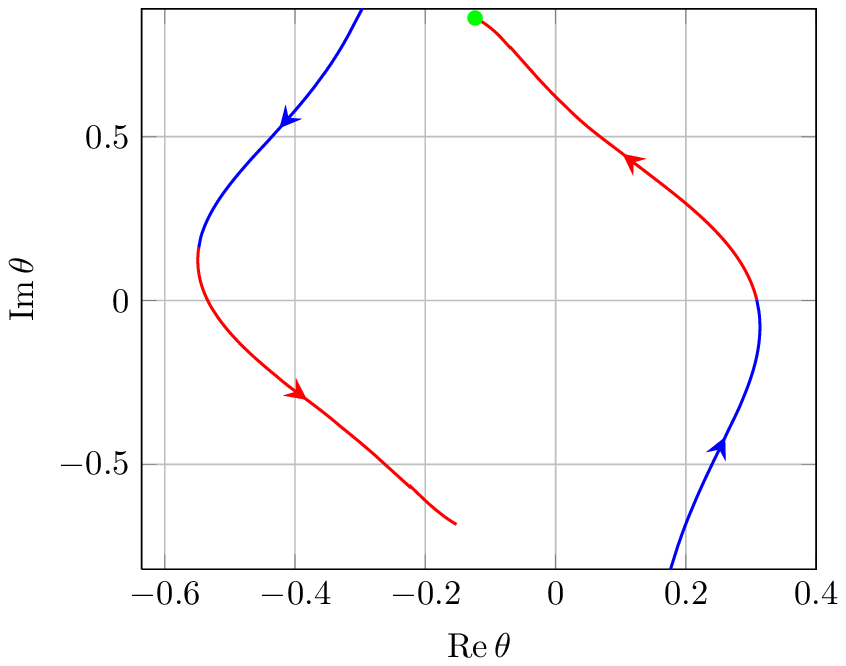}
	\includegraphics[scale=.8]{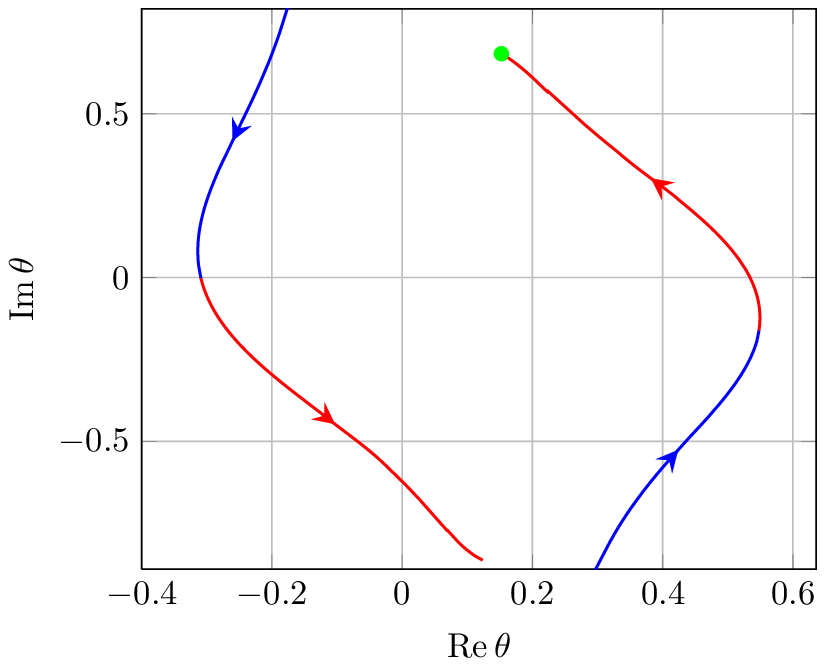}
	\caption{Crossing solutions of the functions $\tilde{\Yf}_{1,2}(\theta)$ (left) and $\tilde{\Yf}_{3,1}(\theta)$ (right) during the continuation Eq.\  (\ref{eq:7pt_mmm}). The mirror symmetry $\tilde{\Yf}_{1,2}(\theta)=\tilde{\Yf}_{3,1}(-\theta)$ is due to our choice $|m_1|=|m_2|=10$, $C_1=-C_2=\mathrm{arccosh}\left(\frac{3}{5}\right)$ at the beginning of the continuation. The arrows indicate the direction of growing $\varphi$. We change colors once the first solution crosses the real axis.}
	\label{fig:7pt_p3_mmm_crossing}
\end{figure}

We then turn to the numerical analysis of $\mathrm{A}_\mathrm{free}$ for this Regge region.
For the first time, we find that four solutions $\tilde{\Yf}_{a,s}(\theta)=-1$ from two different Y-functions, namely $\tilde{\Yf}_{3,1}$ and $\tilde{\Yf}_{1,2}$, cross the real axis.
The endpoints of these crossing solutions will be called $\theta_{12\pm}$ and $\theta_{31\pm}$, respectively.
The corresponding crossing plots are shown in figure \ref{fig:7pt_p3_mmm_crossing}.
From figure \ref{fig:7pt_p3_mmm_crossing} it is obvious that difference of the two pairs of crossed Bethe roots is given by $i\frac{\pi}{2}$.
However, their absolute position does not seem to be a special point in the $\theta$-plane.
This is due to the relatively poor numerical convergence of the Y-functions for this particular path.
In fact, while for all other paths studied so far mass parameters of the size of $\mathcal{O}(10)$ were enough to produce convergent Y-functions after only one iteration of the integral kernels, this is no longer true for this path.
Therefore, the endpoints not ending on a distinguished point in the $\theta$-plane is a reflection of the fact that we are not close enough to the limit $|m_s|\rightarrow \infty$ yet.
Indeed, if we study the endpoint position of the Bethe roots as a function of the initial mass parameter we find the results shown in figure \ref{fig:7pt_pmmm_convergence} as we increase the initial mass parameters towards infinity.
The result of the numerical analysis then is that we have four crossing Bethe roots, two ending at $+i\frac{\pi}{4}$ and two ending at $-i\frac{\pi}{4}$.
\begin{figure}
  \centering
  \includegraphics[scale=1]{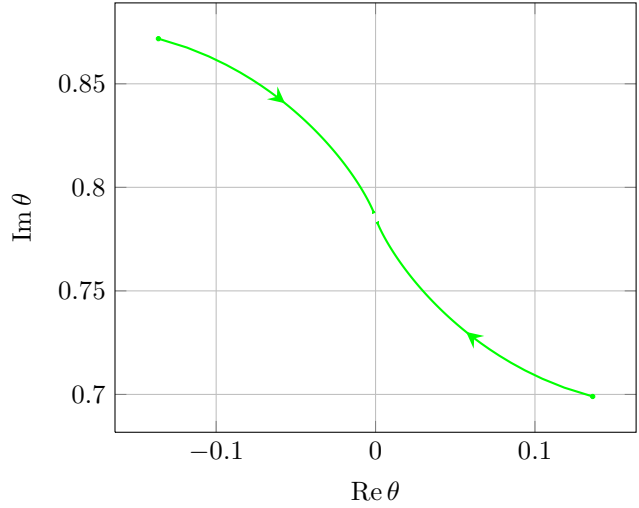}
  \caption{Convergence of the crossing solutions $\theta_{12+}$ and $\theta_{31+}$ against $i\frac{\pi}{4}$. The green dots indicate the position of the green dots in figure \ref{fig:7pt_p3_mmm_crossing} and correspond to an initial value for the mass parameter of $|m_1|=|m_2|=10$. We then increase the initial mass parameter up to $|m_1|=|m_2|=2000$ until the convergence against $i\frac{\pi}{4}$ becomes obvious.}
  \label{fig:7pt_pmmm_convergence}
\end{figure}
This numerical result can be backed up by analytical considerations as for the other paths.
We begin by writing out the Y-system equations at the endpoint of the continuation:
\begin{align}
  \nonumber\log\tilde{\Yf}'_{1,s}(\theta)=&-|m_s|'\cosh\theta-C'_s+\sum\limits_{a',s'}\int d\theta' K_{s,s'}^{1,a'}\left(\theta-\theta'+i\phi'_s-i\phi'_{s'}\right)\log\left(1+\tilde{\Yf}'_{a',s'}(\theta')\right)\\
  &+\log \frac{S_{s,2}^{1,1}(\theta-\theta_{12-}+i\phi'_s-i\phi'_2)}{S_{s,2}^{1,1}(\theta-\theta_{12+}+i\phi'_s-i\phi'_2)}\frac{S_{s,1}^{1,3}(\theta-\theta_{31-}+i\phi'_s-i\phi'_1)}{S_{s,1}^{1,3}(\theta-\theta_{31+}+i\phi'_s-i\phi'_1)},\label{eq:p7_pmmm_ysys1ac}\\
 \nonumber\log\tilde{\Yf}'_{2,s}(\theta)=&-\sqrt{2}|m_s|'\cosh\theta+\sum\limits_{a',s'}\int d\theta' K_{s,s'}^{2,a'}\left(\theta-\theta'+i\phi'_s-i\phi'_{s'}\right)\log\left(1+\tilde{\Yf}'_{a',s'}(\theta')\right)\\
 &+\log \frac{S_{s,2}^{2,1}(\theta-\theta_{12-}+i\phi'_s-i\phi'_2)}{S_{s,2}^{2,1}(\theta-\theta_{12+}+i\phi'_s-i\phi'_2)}\frac{S_{s,1}^{2,3}(\theta-\theta_{31-}+i\phi'_s-i\phi'_1)}{S_{s,1}^{2,3}(\theta-\theta_{31+}+i\phi'_s-i\phi'_1)},\label{eq:p7_pmmm_ysys2ac}
 \end{align}
 \begin{align}
  \nonumber\log\tilde{\Yf}'_{3,s}(\theta)=&-|m_s|'\cosh\theta+C'_s+\sum\limits_{a',s'}\int d\theta' K_{s,s'}^{3,a'}\left(\theta-\theta'+i\phi'_s-i\phi'_{s'}\right)\log\left(1+\tilde{\Yf}'_{a',s'}(\theta')\right)\\
  &+\log \frac{S_{s,2}^{3,1}(\theta-\theta_{12-}+i\phi'_s-i\phi'_2)}{S_{s,2}^{3,1}(\theta-\theta_{12+}+i\phi'_s-i\phi'_2)}\frac{S_{s,1}^{3,3}(\theta-\theta_{31-}+i\phi'_s-i\phi'_1)}{S_{s,1}^{3,3}(\theta-\theta_{31+}+i\phi'_s-i\phi'_1)}.\label{eq:p7_pmmm_ysys3ac}
\end{align}
From the standard two standard endpoint conditions $-1=\tilde{\Yf}'_{1,2}(\theta_{12\pm})=\tilde{\Yf}'_{3,1}(\theta_{31\pm})$, we only learn that
\begin{equation}
  \theta_{12+}-\theta_{12-}=i\frac{\pi}{2}\quad \mathrm{and}\quad \theta_{31+}-\theta_{31-}=i\frac{\pi}{2}.
  \label{eq:endp_simp_pmmm}
\end{equation}
However, we can also take into account the finiteness of ratios of Y-functions at the endpoint of the continuation,
\begin{align}
\nonumber  1&=\frac{\tilde{\Yf}'_{1,2}(\theta_{12-})}{\tilde{\Yf}'_{1,2}(\theta_{12+})}=-e^{|m_2|'(\cosh\theta_{12+}+i\sinh\theta_{12+})}\left(\frac{1+\cosh(\theta_{12+}-\theta_{31+})+i\sinh(\theta_{12+}-\theta_{31+})}{1-\cosh(\theta_{12+}-\theta_{31+})-i\sinh(\theta_{12+}-\theta_{31+})}\right)\\
&=\frac{\tilde{\Yf}'_{3,1}(\theta_{31-})}{\tilde{\Yf}'_{3,1}(\theta_{31+})}=-e^{|m_1|'(\cosh\theta_{31+}+i\sinh\theta_{31+})}\left(\frac{1-\cosh(\theta_{12+}-\theta_{31+})-i\sinh(\theta_{12+}-\theta_{31+})}{1+\cosh(\theta_{12+}-\theta_{31+})+i\sinh(\theta_{12+}-\theta_{31+})}\right),
\label{eq:finite_ratios}
\end{align}
where we have already used Eq.\  (\ref{eq:endp_simp_pmmm}).
As before, the S-matrix factors have to either diverge or go to zero as we send the mass parameters to infinity to give a finite product, depending on the sign of the exponent.
However, we see that in Eq.\  (\ref{eq:finite_ratios}) the S-matrix factors of the two equations are inverses of each other.
Therefore, one of the S-matrix factors has to go to zero, while the other diverges.
This is satisfied if
\begin{equation}
  \theta_{12+}=\theta_{31+}.
  \label{eq:eq_12_31}
\end{equation}
This leaves us with one undetermined endpoint.
Taking the product of the two equations (\ref{eq:finite_ratios}) and using Eq.\  (\ref{eq:eq_12_31}),
\begin{equation}
  1=e^{(|m_1|'+|m_2|')\left(\cosh\theta_{12+}+i\sinh\theta_{12+}\right)},
\end{equation}
we see that the driving term has to be zero which finally gives us $\theta_{12+}=\theta_{31+}=i\frac{\pi}{4}$.
Although certainly more complicated than in the other cases studied so far, this is a very pleasing result, as it reinforces our belief that we only need discrete input from the numerical calculations and can calculate the endpoints analytically.

\subsubsection{Calculation of the remainder function}
We now have all the necessary information to calculate the remainder function $R_{7,---}$.
As the calculation is similar to the one presented in section \ref{sec:region_mmp}, we will be brief.
First, we calculate the cross ratios from Eqs.\ (\ref{eq:p7_pmmm_ysys1ac})-(\ref{eq:p7_pmmm_ysys3ac}) and obtain
\begin{equation}
\begin{alignedat}{3}
  &u_{11}^\prime=1+\varepsilon_2^\prime\left(\frac{1}{w_2^\prime}+\gamma w_2^\prime+2\sqrt{-\gamma}\sinh C_2^\prime\right),\quad &&u_{21}^\prime=-\gamma w_2^\prime \varepsilon_2^\prime,\quad &&u_{31}^\prime=-\frac{\varepsilon_2^\prime}{w_2^\prime},\\
  &u_{12}^\prime=1+\varepsilon_1^\prime\left(w_1^\prime+\frac{\gamma}{w_1^\prime}+2\sqrt{-\gamma}\sinh C_1^\prime\right),\quad && u_{22}^\prime=-w_1^\prime \varepsilon_1^\prime,\quad && u_{31}^\prime=-\gamma\frac{\varepsilon_1^\prime}{w_1^\prime}.
\end{alignedat}
  \label{eq:7pt_pmmm_crs_ac_p3}
\end{equation}
After imposing the identification $u_{as}^\prime=u_{as}$, we find the new parameters to be given by
\begin{equation}
\begin{alignedat}{3}
  &\varepsilon_1^\prime=\frac{\varepsilon_1}{\sqrt{\gamma}},\quad && w_1^\prime=-\sqrt{\gamma}w_1,\quad && \cosh C_1^\prime=-\sinh C_1,\\
  &\varepsilon_2^\prime=\frac{\varepsilon_2}{\sqrt{\gamma}},\quad && w_2^\prime=-\frac{w_2}{\sqrt{\gamma}},\quad && \cosh C_2^\prime= \sinh C_2.
\end{alignedat}
  \label{eq:7pt_pmmm_params_ac}
\end{equation}
Going through the different contributions of the amplitude, we find the following results:
\begin{align}
  A_{\mathrm{free}}^\prime&\cong\sqrt{2}\log\varepsilon_1+\sqrt{2}\log\varepsilon_2-\sqrt{2}\log\gamma,\\
  A_{\mathrm{per}}^\prime-A_{\mathrm{per}}&\cong -\frac{1}{2}\left(i\pi+\log\gamma\right)\log\left(\varepsilon_1\varepsilon_2\right)-\frac{i\pi}{2}\log\left(\frac{w_1}{w_2}\right)+\mathrm{const.}
\end{align}
For $\Delta'$, a small subtlety appears due to the non-trivial rotation of $\tilde{u}$, which gives rise to contributions $\sim\log(1-\tilde{u})$ and $\sim\log\tilde{u}$.
$\tilde{u}$ appears explicitly in the answer.
However, we will not replace it with an expression in terms of the $\varepsilon_s$ and $w_s$, but just use the fact that $1-\tilde{u}\sim \mathcal{O}(\varepsilon^2)$ as can be seen from Eq.\  (\ref{eq:gram_mrl}).
This allows us to drop the term $\sim\log\tilde{u}$ and we obtain
\begin{align}
   \Delta^\prime-\Delta\cong -i\frac{\pi}{2}\log\left(\varepsilon_1\varepsilon_2\right)+\frac{i\pi}{2}\log\left(\frac{w_2}{w_1}\right)-i\frac{\pi}{2}\log\left(1-\tilde{u}\right)+\mathrm{const.}
\end{align}
Putting all results together, and using $\log\gamma=\log\left|\gamma\right|-3i\pi$ we find for the remainder function
\begin{equation}
  R_{7,---}+i\delta_{7,---}=\frac{\sqrt{\lambda}}{2\pi}\left(e_2\log\left(\varepsilon_1\varepsilon_2 \right)+\mathrm{const.}\right),
  \label{eq:rem_fct_p3}
\end{equation}
which is just the sum of the remainder functions for the paths $P_{7,--+}$ and $P_{7,+--}$, presented in section \ref{sec:region_mmp} and appendix \ref{sec:region_pmm}, respectively.
Exponentiating, we find our final result
\begin{equation}
  \left. e^{R_{7,---}+i\delta_{7,---}}\right|_{\mathrm{MRL}}\sim\left((1-u_{11})(1-u_{12})\sqrt{\tilde{u}_{21}\tilde{u}_{31}\tilde{u}_{22}\tilde{u}_{32}}\right)^{\frac{\sqrt{\lambda}}{2\pi}e_2},
  \label{eq:7pt_pmmm_expr}
\end{equation}
where
\begin{equation}
  \delta_{7,---}=\frac{\sqrt{\lambda}}{4}\log\left(\frac{(1-u_{11})(1-u_{12})}{1-\tilde{u}}\sqrt{\tilde{u}_{21}\tilde{u}_{31}\tilde{u}_{22}\tilde{u}_{32}}\right)+\frac{\sqrt{\lambda}}{4}\log\left(\frac{\tilde{u}_{21}}{\tilde{u}_{31}}\frac{\tilde{u}_{32}}{\tilde{u}_{22}}\right).
  \label{eq:phase_pmmm}
\end{equation}
This ends our study of this path and we turn to the last remaining Regge region.
\subsection{Remainder function in the Regge region $(-+-)$}
\label{sec:p7_mpm}
In section \ref{sec:kinematics} we have identified four interesting Regge regions.
The one we have not discussed so far is $P_{7,-+-}$, which we examine using the continuation
\begin{equation}
\begin{alignedat}{4}
& u_{11}(\varphi)=e^{2i\varphi} u_{11},\quad && u_{21}(\varphi)=e^{-i \varphi}u_{21},\quad && u_{31}(\varphi)=e^{i \varphi} u_{31}, && \\
& u_{12}(\varphi)=e^{2i\varphi}u_{12},\quad && u_{22}(\varphi)=e^{i\varphi} u_{22},\quad && u_{32}(\varphi)=e^{-i\varphi}u_{32},\quad && \tilde{u}(\varphi)=e^{-2i\varphi}\tilde{u}.
\end{alignedat}
\label{eq:7pt_pmpm_crs}
\end{equation}
No deformation of paths is needed to satisfy the Gram relation and we can study the crossing solutions as before\footnote{More precisely, there is a deformation of the path of $\tilde{u}$ which satisfies the Gram relation and has the same winding number as the naive path. Since $\tilde{u}$ only appears explicitly in $\Delta$, this deformation is irrelevant.}.
However, for this particular path we see no crossing solutions.
The Y-system equations therefore remain unmodified at the endpoint of the continuation, and the remainder function is trivial up to a phase,
\begin{equation}
  \left.e^{R_{7,-+-}+i\delta_{7,-+-}}\right|_{\mathrm{MRL}}\sim 1,
  \label{eq:7pt_pmpm_rem_fct}
\end{equation}
with
\begin{equation}
  \delta_{7,-+-}=\frac{\sqrt{\lambda}}{4} \log\left(\frac{1}{(1-\tilde{u})}\frac{1}{\tilde{u}_{22}\tilde{u}_{31}}\right),
  \label{eq:7pt_pmpm_phase}
\end{equation}
which arises from the continuation of $\Delta$.
This is a rather surprising result, as we expected to see a contribution of all three cuts, as predicted by the weak coupling analysis (cf. section \ref{sec:wc}).
It may, however, be that our choice of path is too naive for this region.
For example, it is conceivable that this region should rather be probed by first following the path $P_{7,---}$ and only then flipping the middle gluon back up.
Since crossing solutions occur for $P_{7,---}$ it would be interesting to see if those solutions cross back to give the trivial result Eq.\  (\ref{eq:7pt_pmpm_rem_fct}) when flipping the middle gluon back up or if we find a different result.
It should be noted that this issue is not present in the weak coupling description where the various regions can be studied without specifying a path which connects them.

\section{Conclusions and outlook}
In this paper, we have presented a general algorithm for the calculation of $n$-gluon scattering amplitudes in strongly coupled $\mathcal{N}=4$ SYM in various multi-Regge regions. It is based on the central observation that the multi-Regge limit of gauge
theory corresponds to the infrared or large mass limit of the TBA equations that
describe the strong coupling regime. In this limit, the quantum fluctuations of the
one-dimensional auxiliary integrable system can be neglected which leads to drastic
simplifications. In particular, the integral equations that describe excitations of
the one-dimensional system are replaced by a set of algebraic Bethe Ansatz equations
and the energy of excitations is evaluated as the sum of bare energies. These general
observations are highly relevant for the evaluation of Regge cut contributions at
strong coupling since such cut contributions can be argued to be associated with
special excitations of the auxiliary one-dimensional quantum systems. Once the
relevant solution of the Bethe Ansatz equations has been identified, one can build
the associated Y-functions, reconstruct the cross ratios and compute the free
energy and thereby the remainder function at strong coupling.  \par
A crucial ingredient in this program is to decide which solution of the
Bethe Ansatz is actually relevant for any given multi-Regge region. At
the moment we do that by analytically continuing the cross ratios along
a prescribed path and following the solutions of  $\tilde{\Yf}_{a,s}
(\theta)=-1$ numerically. We characterized the paths in section
\ref{sec:mrl_regions}. The main difficulty lies in finding paths which
are compatible with the conformal Gram relations. Constructing these
paths has to be done case-by-case so far and we showed an explicit
example for the $7$-point case. Finding paths for a general number of
gluons would require a better understanding of the conformal Gram
relations for arbitrary $n$. The numerical analysis is explained in
section \ref{sec:gen_alg}. This analysis, too, has to be carried out
for each Regge region separately. We would like to stress again that
the numerical analysis just provides discrete data, namely which
solutions $\tilde{\Yf}_{a,s}(\theta)=-1$ cross the integration contour.
The calculation of the remainder function is therefore not limited by
numerical accuracy.\par
As specific examples of our algorithm, we calculate the $7$-gluon
amplitude in four Regge regions. For the remainder functions
$R_{7,--+}$, $R_{7,+--}$ and $R_{7,---}$ we find that the result
can be expressed only using the functions which appear already in
the $6$-gluon case for the region $R_{6,--}$. This is in remarkable
agreement with the weak-coupling predictions presented in section
\ref{sec:wc} and strengthens the belief that the analytic structure
of the scattering amplitude as predicted by Regge theory is preserved
at any value of the coupling constant.\par
Our findings for the remainder function $R_{7,-+-}$ disagree
with the weak-coupling predictions - while we obtain a trivial remainder function (up to a phase), we expected to find contributions of
three Regge cuts. As stated in section \ref{sec:p7_mpm}, we think
that this mismatch might arise because the path we have chosen is
too naive. It may well be that instead of going to the region
$(-+-)$ immediately, this particular region should rather be probed
by a stepwise continuation, first going to the region $(---)$ and
then flipping the middle gluon back up.
Since we know that crossing solutions occur for the path $P_{7,---}$
it would be interesting to see whether those solutions cross back to
give a trivial remainder function when flipping the middle gluon back
up or whether the remainder function at strong coupling is indeed
sensitive to the way the continuation is performed.
This problem will be addressed in the future.\par
From our results, many ideas for future research emerge.
For example, it would be very interesting to study the $8$-gluon
amplitude. In this case it is expected from the weak-coupling
perspective that in some Regge regions, the functions describing
the $6$-gluon case are no longer sufficient to describe the
remainder function. This is due to the appearance of a new state
which physically corresponds to a bound state of three Reggeons.
It would therefore be relevant to understand if and how this new
state shows up from the strong coupling perspective. Furthermore,
since our $7$-point results can be expressed through the functions
appearing already in the $6$-point amplitude it would be interesting
to see if our results can be related to the conjectured
expressions of \cite{Basso:2014pla}.

\acknowledgments
We would like to thank Benjamin Basso, Simon Caron-Huot, Yasuyuki Hatsuda, Andrey Kormilitzin, Lev Lipatov, Arthur Lipstein and Amit Sever for valuable discussions.
This work was supported by the SFB676.
The research leading to these results has also received funding from
the People Programme (Marie Curie Actions) of the European Union's
Seventh Framework Programme FP7/2007-2013/ under REA Grant Agreement
No 317089 (GATIS).

\begin{appendix}

\section{Conformal Gram relations}
\label{sec:gram}

For a general scattering process involving $n$ particles, there are $3n-10$
independent Mandelstam invariants. As we review in
the main text, we can build many more Mandelstam invariants $p_{i}p_j$ from the $n$
four-momenta of the external gluons. Due to this mismatch, there have to be relations
among these Lorentz invariants. To find additional relations, we have to recall that
in a four-dimensional space there can be at most four linearly independent vectors.
Every additional vector can then be expressed as a linear combination of these four basis vectors.
Without loss of generality, let us choose the momentum vectors $p_1,..., p_4$ to be a basis.
We then have relations
\begin{equation}
\sum\limits_i^{1..4, l} c_i p_i\ =\ 0\ ,
\end{equation}
for coefficients $c_i$ and $l=5,..,n-1$.\footnote{The vector $p_n$ which is not considered
here can always be expressed through the other momenta via overall momentum conservation.}
Multiplication with $2p_j$ gives rise to relations among the Lorentz invariants. The
condition that these relations have non-trivial solutions for the $c_i$ can be cast into
a different form by writing down the $(n-1)\times(n-1)$-matrix of inner products of the
momenta,
\begin{equation}
P \ = \ \begin{pmatrix}
0 & p_1p_2 & p_1p_3 & \cdots & p_1p_{n-1}\\
p_2p_1 & 0 & p_2p_3 & \cdots & p_2p_{n-1}\\
p_3p_1 & p_3p_2 & 0 & \cdots & p_3p_{n-1}\\
\vdots & \vdots & \vdots & \ddots & \vdots \\
p_{n-1}p_1 & p_{n-1}p_2 & p_{n-1}p_3 & \cdots & 0
\end{pmatrix},
\label{eq:coeffmat}
\end{equation}
where the main diagonal contains only zeros, since we are scattering massless gluons.
Since we chose $p_1,...,p_4$ to be a basis, these vectors are linearly independent,
which is equivalent to saying that the upper left $4\times4$-minor
\begin{equation}
\begin{vmatrix}
0 & p_1p_2 & p_1p_3 & p_1p_4\\
p_2p_1 & 0 & p_2p_3 & p_2p_4\\
p_3p_1 & p_3p_2 & 0 & p_3p_4\\
p_4p_1 & p_4p_2 & p_4p_3 & 0
\end{vmatrix}\neq 0
\label{eq:4minor}
\end{equation}
does not vanish. Linear dependence of additional vectors then translates into the
statement that every $5\times5$-minor we can build by adding a column and a row of
Eq.\  (\ref{eq:coeffmat}) to the minor Eq.\  (\ref{eq:4minor}) has to vanish,
\begin{equation}
\begin{vmatrix}
0 & p_1p_2 & p_1p_3 & p_1p_4 & p_1p_k\\
p_2p_1 & 0 & p_2p_3 & p_2p_4 & p_2p_k\\
p_3p_1 & p_3p_2 & 0 & p_3p_4 & p_3p_k\\
p_4p_1 & p_4p_2 & p_4p_3 & 0 & p_4p_k\\
p_lp_1 & p_lp_2 & p_lp_3 & p_lp_4 & p_lp_k\\
\end{vmatrix}=0.
\end{equation}
These relations are called \textit{Gram determinant relations}.
Since the coefficient matrix Eq.\  (\ref{eq:coeffmat}) is symmetric, this gives
rise to $\frac{1}{2}(n-4)(n-5)$ relations.
In total, we then have
\begin{equation}
\frac{n(n-3)}{2}-\frac{(n-5)(n-4)}{2}\ =\ 3n-10
\end{equation}
invariants left, which is exactly the number of independent variables.\par

This, however, is not good enough for our given problem. In the above
description we have only considered Lorentz symmetry of the variables.
However, since the remainder function in $\mathcal{N}=4$ SYM is dual conformal invariant, non-trivial
kinematic information has to be expressed by conformal cross ratios.
We therefore face the problem of writing down Gram relations that obey dual
conformal symmetry. This problem is solved in \cite{Eden:2012tu}.\par
As a first step, we change from momentum variables to the dual variables
$p_i=x_{i-1}-x_{i}$.
Of course, everything we said above still applies to these variables.
We then lift four-dimensional momentum space to a six-dimensional
space on which the four-dimensional dual conformal symmetry $\mathrm{SO}(4,2)$
is realized as rotation symmetry of the null vectors $X_i^\mu:=(1, x_i^2,
x_i^\mu)$. Note that these vectors are denoted in light-cone coordinates,
such that
\begin{equation}
X_i^2=2(1\cdot x_i^2)-2x_i^\mu x_{i\mu}=0,
\end{equation}
as well as
\begin{equation}
X_i\cdot X_j=x_j^2+x_i^2-2x_i\cdot x_j=\left(x_i-x_j\right)^2=x_{i,j}^2.
\end{equation}
Since we are in six-dimensional space, at most six vectors $X_i$ can be
linearly independent. Without loss of generality, we choose $X_1,...,X_6$
as basis vectors. Following the same arguments as above, we find
$\frac{1}{2}(n-5)(n-6)$ conditions of the form
\begin{equation}
\mathrm{Gram}(1,2,3,4,5,6,j,k)=0,
\end{equation}
where the above notation indicates that the $7\times 7$-minor built from the
six basis vectors as well as row $j$ and column $k$ vanishes. These
relations give rise to polynomial equations in the $x^2_{i,j}$ which are dual
conformally invariant by construction. Having found the relations in the
$x^2_{i,j}$ we can use Eqs.\ (\ref{eq:cr1})-(\ref{eq:cr3}) to rewrite the relations in terms
of the cross ratios. For $n=7$ external gluons the only relation reads
\begin{equation}
  a\tilde{u}^2+b\tilde{u}+c=0
  \label{eq:gramfull}
\end{equation}
with lengthy coefficients
\begin{equation}
\begin{aligned}
  a=&u_{11}u_{12}\left(-1+u_{12}u_{21}+u_{11}u_{32}\right),\\
  b=&-\frac{1}{2}+u_{11}+\frac{1}{2}u_{11}u_{12}+u_{12}u_{21}-2u_{11}u_{12}u_{21}-u_{12}^2u_{21}+u_{22}-u_{11}u_{22}\\
  &-2u_{12}u_{21}u_{22}+u_{11}u_{12}u_{21}u_{22}+u_{12}^2u_{21}^2u_{22}-\frac{1}{2}u_{22}u_{31}+u_{12}u_{21}u_{22}u_{31}\\
  &+\frac{1}{2}u_{11}u_{12}u_{21}u_{32}+\frac{1}{2}u_{11}u_{12}u_{21}u_{22}u_{31}u_{32}+(\mathrm{target}\leftrightarrow\mathrm{projectile}),\\
 c=&\frac{1}{2}-u_{11}-u_{21}+u_{11}u_{21}+u_{12}u_{21}-u_{22}+u_{11}u_{22}+u_{21}u_{22}-u_{11}u_{21}u_{22}\\
 &+u_{12}u_{21}u_{22}-u_{12}u_{21}^2u_{22}+\frac{1}{2}u_{22}u_{31}+u_{21}u_{22}u_{31}-2u_{12}u_{21}u_{22}u_{31}\\
 &-u_{21}u_{22}^2u_{31}+u_{12}u_{21}^2u_{22}^2u_{31}+\frac{1}{2}u_{21}u_{32}-u_{11}u_{21}u_{32}-u_{21}u_{22}u_{31}u_{32}\\
  &+u_{11}u_{21}u_{22}u_{31}u_{32}+\frac{1}{2}u_{21}u_{22}^2u_{31}^2u_{32}+\left(\mathrm{target}\leftrightarrow\mathrm{projectile}\right),
\end{aligned}
\end{equation}
where $\left(\mathrm{target}\leftrightarrow\mathrm{projectile}\right)$ means that the same expression after applying a target-projectile transformation (cf.\ Eq.\  (\ref{eq:target_projectile})), including the constant terms, should be added.
This relation simplifies when approaching the multi-Regge limit.
In fact, setting all small cross ratios to zero, $u_{2s}=u_{3s}=0$, in Eq.\  (\ref{eq:gramfull}) we find
\begin{equation}
  0=(\tilde{u}-1)(1-u_{11}-u_{12}+u_{11}u_{12}\tilde{u}).
  \label{eq:gram_mrl}
\end{equation}

\section{The remainder function in the Euclidean regime}
\label{sec:remeuc}

In this appendix we show that the remainder function of the $7$-point amplitude is trivial in the Euclidean region.
To do so, we use Eqs.\ (\ref{eq:crs7pt}) and (\ref{eq:params_eps_w}) to rewrite all contributions of the remainder function in terms of the variables $\varepsilon_s$ and $w_s$.
Let us begin with $\mathrm{A}_{\mathrm{free}}$.
Remember that in the large $|m_s|$-regime the integrals in Eqs.\ (\ref{eq:ysys1})-(\ref{eq:ysys3}) can be neglected, which leads to the following schematic form of the integrals appearing in $\mathrm{A}_{\mathrm{free}}$:
\begin{equation}
  \frac{|m_s|}{2\pi}\int d\theta \cosh\theta\log\left(1+\tilde{\Yf}_{a,s}(\theta)\right)\approx \frac{|m_s|}{2\pi}\int d\theta \cosh\,\theta e^{-|m_s|\cosh\theta}=\frac{|m_s|}{\pi} K_1(|m_s|),
  \label{eq:afreeeuc}
\end{equation}
where $K_1(x)$ is a modified Bessel function of the second kind \cite{Abramowitz1972}.
Using its large $x$-asymptotics,
\begin{equation}
  xK_1(x)\sim \sqrt{\frac{\pi x}{2}}e^{-x},
\end{equation}
as well as
\begin{equation}
	|m_s|=\left(\log^2\varepsilon_s+\log^2 w_s\right)^\frac{1}{2}\approx-\log\varepsilon_s,
\end{equation}
we see that the above integral Eq.\  (\ref{eq:afreeeuc}) is of $\mathcal{O}(\varepsilon\log\varepsilon)$ and therefore vanishes in the $\varepsilon\rightarrow 0$ limit.
Hence, $\mathrm{A}_{\mathrm{free}}$ does not contribute in this limit.
Next we study $\mathrm{A}_{\mathrm{per}}$.
Starting from Eq.\  (\ref{eq:aper}) and using Eq.\  (\ref{eq:params_eps_w}), we find that
\begin{equation}
\begin{aligned}
\nonumber  \mathrm{A}_{\mathrm{per}}=&\frac{1}{2}\left(\log^2\varepsilon_1+\log^2 w_1+\log^2\varepsilon_2+\log^2 w_2\right.\\
&\quad\left.+\log\varepsilon_1\log\varepsilon_2+\log w_1\log w_2+\log\varepsilon_2\log w_1-\log\varepsilon_1\log w_2\right).
\end{aligned}
  \label{eq:apereuc}
\end{equation}
For the last remaining term of the remainder function, $\mathrm{A}_{\mathrm{BDS-like}}-\mathrm{A}_{\mathrm{BDS}}$, we use Eq.\  (\ref{eq:crs7pt}) in Eq.\  (\ref{eq:deltabc}), expand in $\varepsilon_s$ and keep only the leading terms, finding
\begin{equation}
\begin{aligned}
\Delta=&-\frac{\pi^2}{6}-\frac{1}{2}\left(\log^2\varepsilon_1+\log^2 w_1+\log^2\varepsilon_2+\log^2 w_2\right)\\
&-\frac{1}{2}\left(\log\varepsilon_1\log\varepsilon_2-\log\varepsilon_1\log w_2+\log\varepsilon_2\log w_1+\log w_1\log w_2\right).
\end{aligned}
\label{eq:deltaeuc}
\end{equation}
Summing all terms, we find that only a constant remainder function remains, as it must.
It should be noted that this constant, $-\frac{\pi^2}{6}$, comes solely from the leading term in the expansion of the $\mathrm{Li}_2(1-u_{i})$, which occur with opposite sign in $\mathrm{A}_{\mathrm{BDS}}$.
Hence, this constant will cancel in the full amplitude.

\section{Kernels of the rewritten Y-system}
\label{sec:app_kernels}
In this appendix we spell out the Y-system kernels $\mathcal{K}_{s,s'}^{a,a'}$ of the rewritten Y-system Eqs.\  (\ref{eq:newysys1})-(\ref{eq:newysys3}), which we use in the numerical analysis of $\mathrm{A}_\mathrm{free}$.
$K_i(x)$ denote the kernels of the standard Y-system (cf.\ Eq.\  (\ref{eq:kernels})).
In the following, $s\pm 1$ denotes the unique possible choice for $s'$ in a given kernel.
Furthermore, if a formula holds for both $a'=1$ and $a'=3$ we just write $a'=2\pm 1$.
\begin{alignat*}{2}
  &\mathcal{K}_{s,s}^{1,2\pm 1}(\theta,\theta')&&=K_1(\theta')\cosh\theta-K_1(\theta-\theta')\\
  &\mathcal{K}_{s,s}^{1,2}(\theta,\theta')&&=K_2(\theta')\cosh\theta-K_2(\theta-\theta')\\
  &\mathcal{K}_{s,s\pm1}^{1,2}(\theta,\theta')&&=-K_1(-\theta'+i\phi_s-i\phi_{s\pm1})\cosh\theta+K_1(\theta-\theta'+i\phi_s-i\phi_{s\pm1})\\
  &\mathcal{K}_{s,s\pm1}^{1,1}(\theta,\theta')&&=\frac{1}{2}\left(K_2(\theta-\theta'+i\phi_s-i\phi_{s\pm1})-K_2(-\theta'+i\phi_s-i\phi_{s\pm1})\cosh\theta\right)\\
  & &&\quad+\frac{1}{2}(-1)^{s+1}\left(K_3(\theta-\theta'+i\phi_s-i\phi_{s\pm1})-K_3(-\theta'+i\phi_s-i\phi_{s\pm1})\right)\\
  &\mathcal{K}_{s,s\pm1}^{1,3}(\theta,\theta')&&=\frac{1}{2}\left(K_2(\theta-\theta'+i\phi_s-i\phi_{s\pm1})-K_2(-\theta'+i\phi_s-i\phi_{s\pm1})\cosh\theta\right)\\
  & &&\quad+\frac{1}{2}(-1)^{s}\left(K_3(\theta-\theta'+i\phi_s-i\phi_{s\pm1})-K_3(-\theta'+i\phi_s-i\phi_{s\pm1})\right)\\
  &\mathcal{K}_{s,s}^{2,2\pm 1}(\theta,\theta')&&=\sqrt{2}K_1(\theta')\cosh\theta-K_2(\theta-\theta')\\
  &\mathcal{K}_{s,s}^{2,2}(\theta,\theta')&&=\sqrt{2}K_2(\theta')\cosh\theta-2K_1(\theta-\theta')\\
  &\mathcal{K}_{s,s\pm1}^{2,2}(\theta,\theta')&&=-\sqrt{2}K_1(-\theta'+i\phi_s-i\phi_{s\pm1})\cosh\theta+K_2(\theta-\theta'+i\phi_s-i\phi_{s\pm1})\\
  &\mathcal{K}_{s,s\pm1}^{2,2\pm 1}(\theta,\theta')&&=-\frac{1}{\sqrt{2}}K_2(-\theta'+i\phi_s-i\phi_{s\pm1})\cosh\theta+K_1(\theta-\theta'+i\phi_s-i\phi_{s\pm1})\\
  &\mathcal{K}_{s,s}^{3,2\pm 1}(\theta,\theta')&&=K_1(\theta')\cosh\theta-K_1(\theta-\theta')\\
  &\mathcal{K}_{s,s}^{3,2}(\theta,\theta')&&=K_2(\theta')\cosh\theta-K_2(\theta-\theta')\\
  &\mathcal{K}_{s,s\pm1}^{3,2}(\theta,\theta')&&=-K_1(-\theta'+i\phi_s-i\phi_{s\pm1})\cosh\theta+K_1(\theta-\theta'+i\phi_s-i\phi_{s\pm1})\\
  &\mathcal{K}_{s,s\pm1}^{3,1}(\theta,\theta')&&=\frac{1}{2}\left(K_2(\theta-\theta'+i\phi_s-i\phi_{s\pm1})-K_2(-\theta'+i\phi_s-i\phi_{s\pm1})\cosh\theta\right)\\
  & &&\quad+\frac{1}{2}(-1)^{s}\left(K_3(\theta-\theta'+i\phi_s-i\phi_{s\pm1})-K_3(-\theta'+i\phi_s-i\phi_{s\pm1})\right)\\
  &\mathcal{K}_{s,s\pm1}^{3,3}(\theta,\theta')&&=\frac{1}{2}\left(K_2(\theta-\theta'+i\phi_s-i\phi_{s\pm1})-K_2(-\theta'+i\phi_s-i\phi_{s\pm1})\cosh\theta\right)\\
  & &&\quad+\frac{1}{2}(-1)^{s+1}\left(K_3(\theta-\theta'+i\phi_s-i\phi_{s\pm1})-K_3(-\theta'+i\phi_s-i\phi_{s\pm1})\right)
\end{alignat*}

\section{S-matrices}
\label{sec:app_smatrices}
For the new set of kernels written out in appendix \ref{sec:app_kernels}, we have to determine the corresponding S-matrices.
However, this is simple because the new kernels are linear combinations of the original kernels, possibly with prefactors.
Therefore, the new S-matrices too should be given in terms of the original S-matrices Eqs.\ (\ref{eq:basic_smat}).
Recall the original definition of the S-matrices,
\begin{equation}
  -2\pi i K(\theta)=:\partial_{\theta}\log S(\theta).
\end{equation}
Since the new kernels are now in general functions of both $\theta$, $\theta'$ and not just their difference, we have to extend this definition in the following way:
\begin{equation}
  -2\pi i K(\theta,\theta')=:\partial_{\theta'}\log S(\theta,\theta')
  \label{eq:defsmat2}
\end{equation}
This allows us to pick up the residue contributions in a very simple way, as before.
Using this definition, we see that for a kernel of the schematic form
\begin{equation}
  K(\theta,\theta')=f(\theta)K_1(\theta,\theta')+g(\theta)K_2(\theta,\theta')+\dots
  \label{eq:kergenform}
\end{equation}
the S-matrix is given by
\begin{equation}
  S(\theta, \theta')=S_1(\theta,\theta')^{f(\theta)}\cdot S_2(\theta,\theta')^{g(\theta)}\cdot\dots
  \label{eq:smatgenform}
\end{equation}
As a specific example, we spell out the S-matrices for a crossed solution of $\tilde{\Yf}_{3,1}$ in the rewritten Y-system below:
\begin{equation}
\begin{aligned}
  S_{1,1}^{2,3}(\theta,\theta')&=S_1(\theta')^{\sqrt{2}\cosh\theta}S_2(\theta-\theta')\\
  S_{1,1}^{2\pm 1,3}(\theta,\theta')&=S_1(\theta')^{\cosh\theta}S_1(\theta-\theta')\\
  S_{2,1}^{1,3}(\theta,\theta')&=\frac{S_2(-\theta'+i\phi_2-i\phi_1)^{\frac{1}{2}\cosh\theta}S_3(-\theta'+i\phi_2-i\phi_1)^{\frac{1}{2}}}{S_2(\theta-\theta'+i\phi_2-i\phi_1)^{\frac{1}{2}}S_3(\theta-\theta'+i\phi_2-i\phi_1)^{\frac{1}{2}}}\\
  S_{2,1}^{2,3}(\theta,\theta')&=S_2(\theta'-i\phi_2+i\phi_1)^{-\frac{1}{\sqrt{2}}\cosh\theta}S_1(\theta-\theta'+i\phi_2-i\phi_1)^{-1}\\
  S_{2,1}^{3,3}(\theta,\theta')&=\frac{S_2(-\theta'+i\phi_2-i\phi_1)^{\frac{1}{2}\cosh\theta}S_3(\theta-\theta'+i\phi_2-i\phi_1)^{\frac{1}{2}}}{S_2(\theta-\theta'+i\phi_2-i\phi_1)^{\frac{1}{2}}S_3(-\theta'+i\phi_2-i\phi_1)^{\frac{1}{2}}}
\end{aligned}
\end{equation}
Similarly, we obtain all the other S-matrices from the kernels of appendix \ref{sec:app_kernels}.
We refrain from spelling them out here explicitly.

\section{Remainder function in the Regge region $(+--)$}
\label{sec:region_pmm}
\begin{figure}[tb]
  \centering
  \includegraphics[scale=.9]{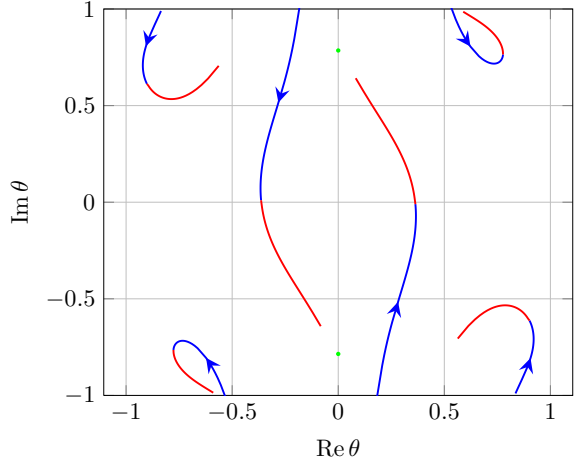}
  \caption{Paths followed by the solutions of $\tilde{\Yf}_{3,1}(\theta)=-1$ during the continuation Eq.\  (\ref{eq:7pt_pmm}). We find that one pair of solutions crosses the real axis. The direction of movement is indicated by arrows on the plot. We change the color of the curve when the pair of solutions crosses the integration contour.}
  \label{fig:7pt_p5_crossing}
\end{figure}

In this section we will briefly show the calculation for the path $P_{7,+--}$ of the cross ratios that was defined in Eq.\ \eqref{eq:7pt_pmm}.
The calculation proceeds along the same steps as the one in section \ref{sec:calcamp}.
Therefore, we only highlight the differences in the following.
For the numerical analysis, we again have to determine the paths of the variables Eq.\  (\ref{eq:newvar}) under the continuation Eq.\  (\ref{eq:7pt_pmm}).
We find the results depicted in figure \ref{fig:P5DrvTerms}.
\begin{figure}[tb]
  \centering
  \includegraphics[scale=.67]{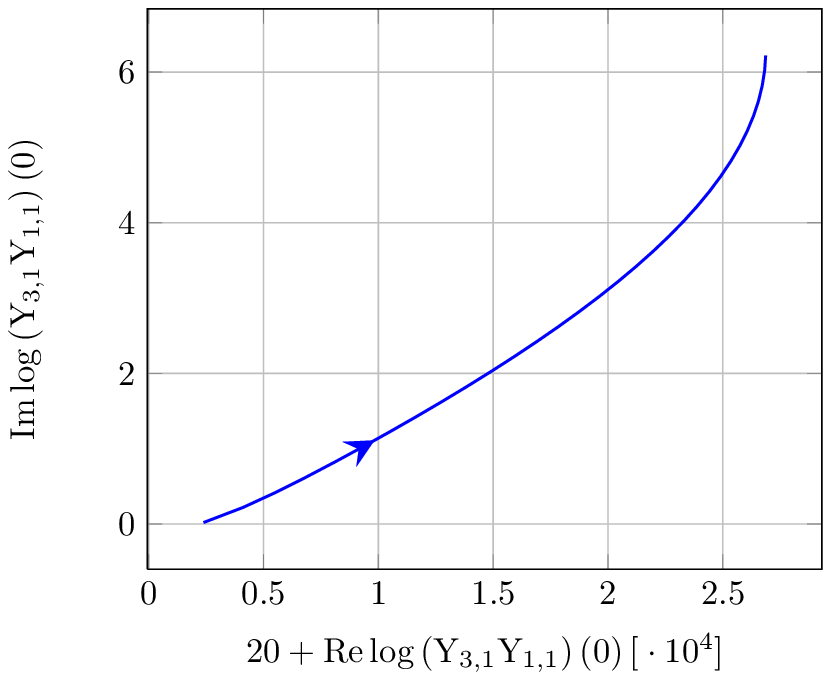}
  \includegraphics[scale=.67]{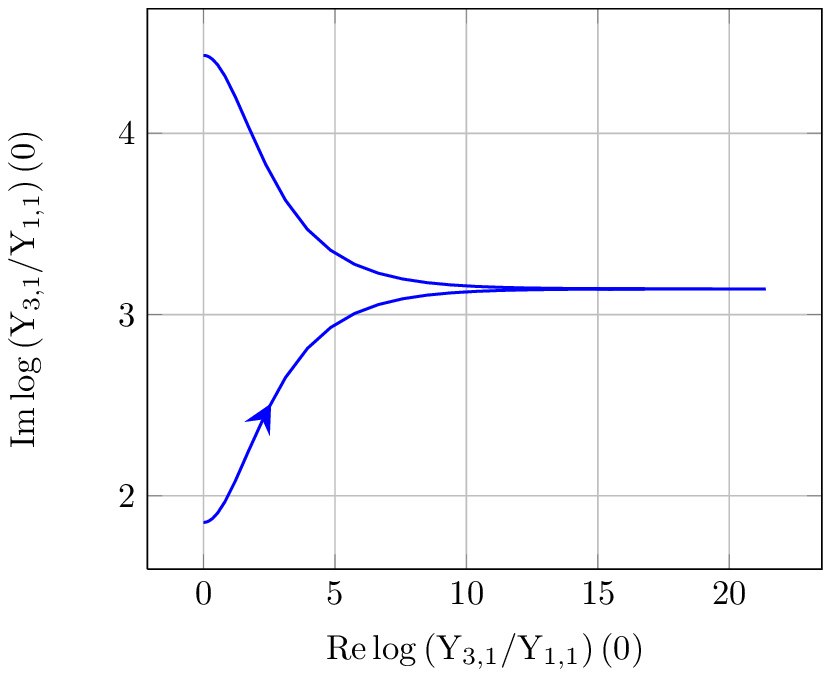}
  \includegraphics[scale=.67]{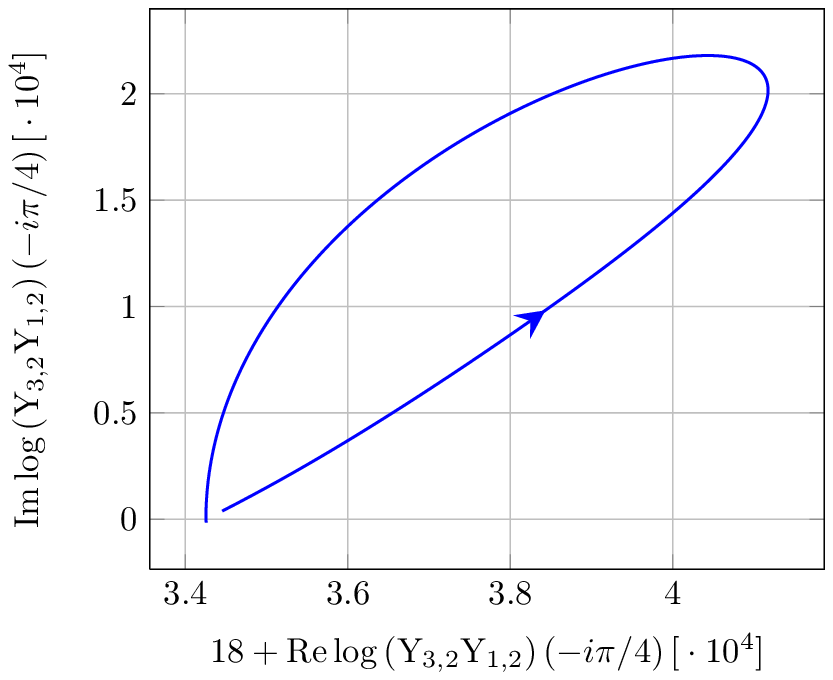}
  \includegraphics[scale=.67]{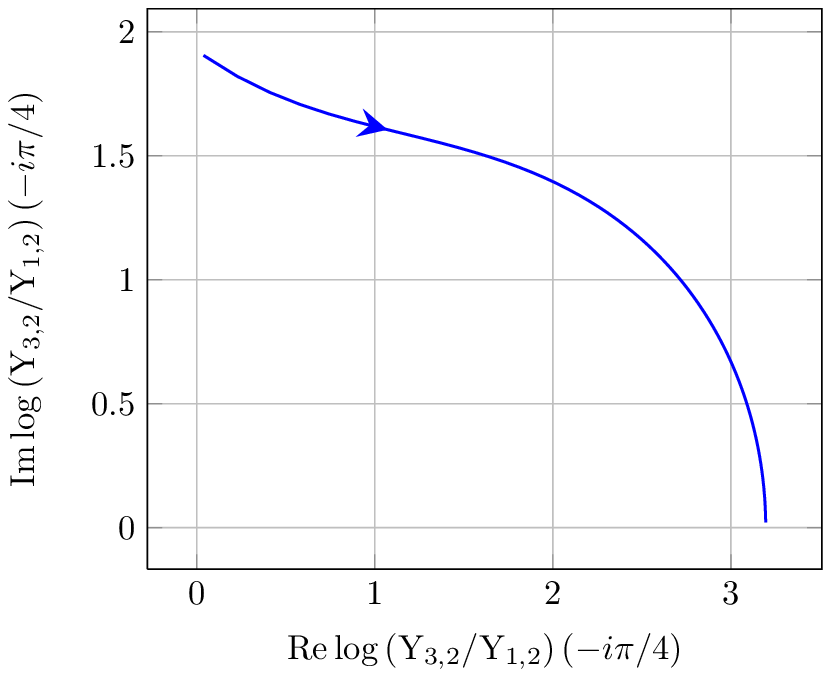}
  \caption{Continuation of the driving terms for the path Eq.\  (\ref{eq:7pt_pmm}). Note that some axes have been rescaled and shifted. The direction of growing $\varphi$ is indicated by the arrows. The plots shown correspond to the parameter choice $|m_1|=10$, $|m_2|=9$, $C_1=\mathrm{arccosh}\left(\frac{3}{5}\right)$, $C_2=\mathrm{arccosh}\left(\frac{4}{7}\right)$ at the starting point.}
  \label{fig:P5DrvTerms}
\end{figure}
Having found the paths of the driving terms, we can follow the solutions $\tilde{\Yf}_{a,s}=-1$ during the continuation, leading to the results shown in figure \ref{fig:7pt_p5_crossing}.
We see that for the path $P_{7,+--}$, two solutions of $\tilde{\Yf}_{3,1}$ cross the real axis and approach $\pm i \frac{\pi}{4}$, which can again be confirmed by an analytic argument using the endpoint conditions Eq.\  (\ref{TBABA}).
Furthermore, we find a pair of solutions of $\tilde{\Yf}_{2,1}$ which approaches the real axis but never crosses.
As before, these solutions do not contribute to the remainder function.
Thus, after neglecting the integrals, the Y-system after the continuation is given by
\begin{align}
  \Yf'_{1,s}&=\left(e^{-|m_s|'\cosh(\theta-i\phi'_s)-C'_s}\right)\frac{S_{s,1}^{1,3}(\theta+i\frac{\pi}{4}-i\phi'_1)}{S_{s,1}^{1,3}(\theta-i\frac{\pi}{4}-i\phi'_1)},\\
  \Yf'_{2,s}&=\left(e^{-\sqrt{2}|m_s|'\cosh(\theta-i\phi'_s)}\right)\frac{S_{s,1}^{2,3}(\theta+i\frac{\pi}{4}-i\phi'_1)}{S_{s,1}^{2,3}(\theta-i\frac{\pi}{4}-i\phi'_1)},\\
  \Yf'_{3,s}&=\left(e^{-|m_s|'\cosh(\theta-i\phi'_s)+C'_s}\right)\frac{S_{s,1}^{3,3}(\theta+i\frac{\pi}{4}-i\phi'_1)}{S_{s,1}^{3,3}(\theta-i\frac{\pi}{4}-i\phi'_1)},
\end{align}
with the S-matrices spelled out in appendix \ref{sec:app_smatrices}.
With this Y-system we determine the cross ratios after continuation to be
\begin{equation}
\begin{alignedat}{3}
  \nonumber u'_{11}&=1+\varepsilon'_2\left(w'_2+\frac{1}{\gamma w'_2}+2\frac{1}{\sqrt{-\gamma}}\sinh C'_2\right),\quad && u'_{21}=-w'_2\varepsilon'_2,\quad && u'_{31}=-\frac{1}{\gamma}\frac{\varepsilon'_2}{w'_2},\\
 u'_{12}&=1-\gamma\varepsilon'_1\left(w'_1+\frac{1}{w'_1}-2\cosh C'_1\right),\quad && u'_{22}=\gamma w'_1\varepsilon'_1,\quad && u'_{32}=\gamma\frac{\varepsilon'_1}{w'_1}.
 \end{alignedat}
 \label{eq:7pt_ppmm_crs}
\end{equation}
Setting $u'_{21}=-u_{21}$, $u'_{31}=-u_{31}$ and $u'_{as}=u_{as}$ for the remaining cross ratios we find
\begin{equation}
\begin{alignedat}{3}
\varepsilon'_1&=\frac{1}{\gamma}\varepsilon_1,\quad && w'_1=w_1,\quad && \cosh C'_1=-\cosh C_1,\\
\varepsilon'_2&=\sqrt{\gamma}\varepsilon_2,\quad && w'_2=\frac{1}{\sqrt{\gamma}}w_2,\quad && \cosh C'_2=\sqrt{1-\left(w_2+\frac{1}{w_2}+\cosh C_2\right)^2}
\end{alignedat}
\label{eq:epsacp2}
\end{equation}
for the analytically continued auxiliary parameters.
The rest of the calculation goes through as for the path $P_{7,--+}$ and we find the result
\begin{equation}
  \left.  e^{\mathrm{R_{7,+--}}+i\delta_{7,+--}}\right|_{\mathrm{MRL}}\sim \left(-(1-u_{12})\sqrt{\tilde{u}_{22}\tilde{u}_{32}}\right)^{\frac{\sqrt{\lambda}}{2\pi}e_2}
  \label{eq:R7_ppmm}
\end{equation}
where
\begin{equation}
  \delta_{7,+--}=\frac{\sqrt{\lambda}}{4}\log\left(\sqrt{\tilde{u}_{22}\tilde{u}_{32}}\right).
  \label{eq:conf_phase_ppmm}
\end{equation}
This result is consistent with the target-projectile symmetry which relates paths $P_{7,--+}$ and $P_{7,+--}$.
\end{appendix}

\end{document}